\documentclass[aps,prb,twocolumn,superscriptaddress]{revtex4-2}
\usepackage[dvips]{graphicx}
\usepackage{comment}
\usepackage{color}
\usepackage{amsmath}
\usepackage{txfonts}
\usepackage{bm}
\usepackage{here}

\begin{document}
\def\val{3}

\title{Pressure-Tuned Metamagnetism and Emergent Three-Body Interactions in CsFeCl$_3$}


\author{K. Nihongi}
\author{T. Kida}
\author{Y. Narumi}
\affiliation{Center for Advanced High Magnetic Field Science (AHMF), Graduate School of Science, The University of Osaka, Toyonaka, Osaka 560-0043, Japan}

\author{Y. Etoh}
\affiliation{Department of Physics, Nihon University, Tokyo 156-8550, Japan}

\author{D. Yamamoto}
\affiliation{Department of Physics, Nihon University, Tokyo 156-8550, Japan}
\affiliation{RIKEN Center for Quantum Computing (RQC), Wako, Saitama 351-0198, Japan}

\author{M. Matsumoto}
\affiliation{Department of Physics, Shizuoka University, Shizuoka 422-8529, Japan}

\author{N. Kurita}
\author{H. Tanaka}
\affiliation{Department of Physics, Tokyo Institute of Technology, Meguro-ku, Tokyo 152-8551, Japan}

\author{K. Yu. Povarov}
\author{S. A. Zvyagin}
\affiliation{Dresden High Magnetic Field Laboratory (HLD-EMFL) and W\"urzburg-Dresden Cluster of Excellence ct.qmat, Helmholtz-Zentrum Dresden-Rossendorf (HZDR), 01328 Dresden, Germany}

\author{J. Wosnitza}
\affiliation{Dresden High Magnetic Field Laboratory (HLD-EMFL) and W\"urzburg-Dresden Cluster of Excellence ct.qmat, Helmholtz-Zentrum Dresden-Rossendorf (HZDR), 01328 Dresden, Germany}
\affiliation{Institut f\"{u}r Festk\"{o}rper- und Materialphysik, TU Dresden, 01062 Dresden, Germany}

\author{K. Kindo}
\author{Y. Uwatoko}
\affiliation{The Institute for Solid State Physics, The University of Tokyo, Kashiwa, Chiba 277-8581, Japan}

\author{M. Hagiwara} \thanks{Corresponding author}
\affiliation{Center for Advanced High Magnetic Field Science (AHMF), Graduate School of Science, The University of Osaka, Toyonaka, Osaka 560-0043, Japan}
\email[e-mail:]{hagiwara@ahmf.sci.osaka-u.ac.jp}



\date{\today}

\begin{abstract}

We present a combined experimental and theoretical study of the triangular-lattice quantum antiferromagnet CsFeCl$_3$ under high magnetic fields and high pressure. Pulsed-field magnetization for the magnetic field along the symmetric $c$ direction at ambient pressure reveals a magnetization process from a nonmagnetic singlet ground state with a nearly linear increase between 3.7 and 10.7\,T, a plateau-like region, and then a sharp stepwise metamagnetic transition near 32\,T. Wide frequency--field range electron spin resonance indicates that the low-field regime originates from the $J = 1$ manifold, while the high-field metamagnetic transition suggests a level crossing between the $J = 1$ and $J = 2$ lowest states. Pulsed-field magnetic susceptibilities measured with a proximity detector oscillator under high pressure show that the low-field nonmagnetic singlet phase is gradually suppressed, while the high-field metamagnetic transition evolves into an increasingly rich pattern of fractional steps. While the observations at low to intermediate fields can be understood within the established spin-1 description, the high-field regime requires a new perspective, which we provide through a projected spin-1/2 framework built from Zeeman-selected crystal-field states not related by time reversal. This construction naturally allows emergent three-body interactions on triangular plaquettes and explains the asymmetric evolution of the fractional steps in the magnetization. Our findings reveal that high-field effective spin models in quantum magnets with separated yet accessible crystal-field multiplets are not constrained to even-body couplings, but can naturally host odd-body terms, opening a broader avenue for realizing field-asymmetric magnetization processes and exotic phases beyond conventional even-body physics.
\end{abstract}

\maketitle

\section{Introduction}

Symmetry principles are central to modern physics, offering a unifying framework for understanding fundamental laws and classifying phases of matter. Among them, time-reversal symmetry (TRS) plays a crucial role across scales, from classical particle motion to coherence phenomena in quantum systems. In relativistic quantum field theory, TRS is embedded within the CPT theorem, which states that the combined operation of charge conjugation (C), parity inversion (P), and time reversal (T) is an exact symmetry of any local, Lorentz-invariant theory~\cite{Lee1981}. TRS is indeed violated in particle physics, most notably in conjunction with CP-violating processes in the weak interaction~\cite{PhysRevLett.13.138, PhysRevLett.109.211801}. In condensed-matter systems, however, the absence of strict Lorentz invariance allows for a wider range of TRS breaking mechanisms. These include both explicit and spontaneous cases, typically arising not from the microscopic Hamiltonian itself, but from external symmetry-breaking fields or emergent order parameters. This leads to a rich variety of phenomena such as magnetism, topological phases, and nonreciprocal transport~\cite{RevModPhys.82.3045, RevModPhys.82.1539}.

In spin systems, TRS breaking is often associated with spontaneous magnetic ordering or the application of an explicit external magnetic field. Nevertheless, many spin Hamiltonians, including the prototypical Heisenberg model and its common extensions, remain invariant under the reversal of all spin directions \( \bm{S}_i \to -\bm{S}_i \) at each site \( i \). This invariance stems from the fact that most typical spin-interaction terms in microscopic Hamiltonians involve even powers of spin operators~\cite{RevModPhys.95.035004,PhysRevB.93.214431}. For example, the Heisenberg exchange \( \bm{S}_i \cdot \bm{S}_j \), the Dzyaloshinskii-Moriya interaction \( \bm{D} \cdot (\bm{S}_i \times \bm{S}_j) \), and the single-ion anisotropy \( (S_i^z)^2 \) are all quadratic in spin operators and remain unchanged under \( \bm{S}_i \to -\bm{S}_i \). {This holds even for more exotic forms of spin interactions, such as the Kitaev~\cite{Takagi2019} and $\Gamma$-type exchanges~\cite{Winter2017}}. Similarly, four-spin interactions such as biquadratic~\cite{PhysRevLett.59.799,PhysRevLett.83.4176} and ring-exchange~\cite{PhysRevLett.94.197202} terms also preserve this inversion symmetry. As a result, although the external magnetic field $\bm{H}$ itself breaks TRS through the linear Zeeman term \( \propto \bm{H} \cdot\sum_i  \bm{S}_i \), the full Hamiltonian usually satisfies the relation \( \mathcal{H}(\{\bm{S}_i\}, \bm{H}) = \mathcal{H}(\{-\bm{S}_i\}, -\bm{H}) \). This combined symmetry leads directly to the observable relation $M(H) = -M(-H)$ for the magnetization $M$ as a function of field $H$, i.e., the curve is an odd function of field, which serves as a key constraint in modeling conventional spin systems.

\begin{figure}[tb]
\begin{center}
\includegraphics[keepaspectratio, scale=0.7]{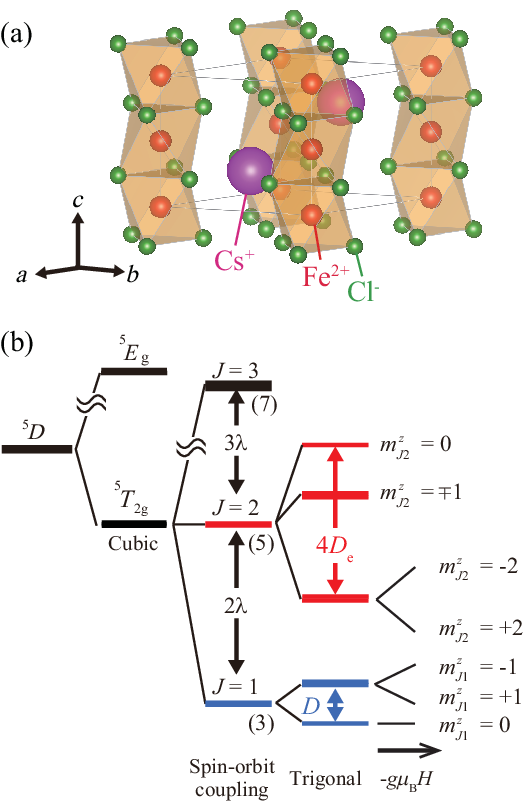}
\caption{ (a) Crystal structure of CsFeCl$_3$ depicted using VESTA~\cite{momma2011vesta}. (b) Schematic view of the energy levels of the Fe$^{2+}$ ion~\cite{hori1989magnetization}.}\label{f1}
\end{center}
\end{figure}

In this work, we explore a case, where the conventional symmetry \( \mathcal{H}(\{\bm{S}_i\}, H) = \mathcal{H}(\{-\bm{S}_i\}, -H) \) is effectively broken in the low-energy description, even though the underlying atomic system is time-reversal symmetric. Specifically, we focus on the high-field regime of the triangular-lattice compound CsFeCl$_3$, which belongs to a hexagonal structure with space group $P6_3/mmc$ and the lattice constants are $a$ = 7.235 {\AA} and $c$ = 6.050 {\AA} at room temperature [Fig.~\ref{f1}(a)]~\cite{Seifert}. Fe$^{2+}$ ions are surrounded by six Cl$^-$ ions forming an octahedron and a one-dimensional chain structure along the $c$-axis. These chains form a triangular lattice in the $ab$-plane. The magnetic Fe$^{2+}$ ion possesses the atomic configuration 3$d^6$, which is a $^5D$ state with an orbital angular momentum $L$ = 2 and a spin momentum $S$ = 2. The $^5D$ state splits into the higher orbital doublet $^5E_g$ and lower orbital triplet $^5T_{2g}$ by the cubic crystalline field as shown in Fig. \ref{f1}(b). The $^5T_{2g}$ state is given by the effective orbital angular momentum $l$ = 1 and $S$ = 2, and split into three multiplets with total angular momentum $J$ = 1, 2, and 3 by spin-orbit coupling and the trigonal distortion of the FeCl$_6$ octahedron. The corresponding single-ion Hamiltonian in a magnetic field along the $z$ axis is given by~\cite{suzuki1981theoretical}
\begin{eqnarray}
{\mathcal{H}_{\rm ion}} = -k\lambda \mbox{\boldmath $l$}\cdot \mbox{\boldmath $S$} - \delta \left((l^{z})^2-\frac{2}{3}\right)-\mu_{\rm B}H(-kl^{z}+2S^{z}), \label{crystal_field_Hamiltonian}
\end{eqnarray}
\noindent
where $k$ and $\lambda$ are the orbital reduction factor $\sim$0.9 and the spin-orbit coupling constant $\sim$3.1 THz, respectively~\cite{euler1978quadrupole}. $\mbox{\boldmath $l$}$ is antiparallel to the real orbital angular momentum $\mbox{\boldmath $L$}$ (= $-k\mbox{\boldmath $l$}$). $\delta$ is the magnitude of the trigonal distortion. The last term is the Zeeman energy and $\mu_{\rm B}$ is the Bohr magneton. Since the FeCl$_6$ octahedron is trigonally elongated along the $c$ axis, the $J$ = 1 state splits into the lowest singlet state ($m_{J1}^z$ = 0) and the upper doublet state ($m_{J1}^z$ = $\pm$1), whereas the $J$ = 2 state splits into the three states ($m_{J2}^z$ = 0, $\pm$1, and $\pm$2). These energy differences can be described by the single-ion anisotropy ($D$ and $D_{\rm e}$).

\begin{figure}[tb]
\begin{center}
\includegraphics[keepaspectratio, scale=0.7]{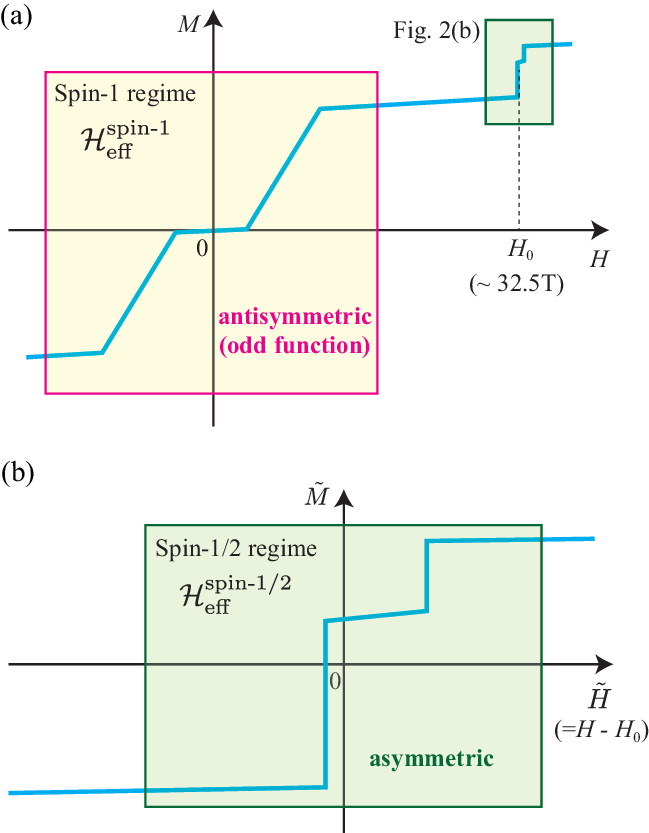}
\caption{(a) Sketch of the magnetization curve of CsFeCl$_3$ for fields applied along the $c$-axis at ambient pressure~\cite{CHIBA1987427,tsuboi1988magnetization}. (b) Enlarged view of the high-field regime. Here, $\tilde{M}$ and $\tilde{H}$ denote the rescaled magnetization and effective field defined within the spin-1/2 model description, which are shifted from the physical magnetization and field origin of the full system.}\label{sketch}
\end{center}
\end{figure}

Reflecting the fact that the lowest single-ion energy level is the \( m_{J1}^z = 0 \) singlet state, CsFeCl$_3$ exhibits a non-magnetic disordered ground state at low temperatures. As schematically shown in Fig.~\ref{sketch}(a), applying a magnetic field along the $c$-axis ($z$-axis) induces a quantum phase transition at around \( H\sim4 \)~T into a magnetically ordered phase, interpreted as a Bose-Einstein condensation (BEC) of magnons~\cite{PhysRevLett.84.5868,PhysRevB.71.224426,Giamarchi2008,RevModPhys.86.563}. The magnetization then increases nearly linearly and approaches a saturation-like behavior around \( H \sim 10 \)~T. This low- to intermediate-field behavior is well described by a spin-1 model involving the three lowest crystal-field levels \( m_{J1}^z = +1, 0, -1 \), as previously established~\cite{yoshizawa1980neutron,kurita2016magnetic,hayashida2019novel,hayashida2018}. However, at higher fields, the system exhibits an unexpected metamagnetic response characterized by sharp steps and quantized magnetization plateaus~\cite{CHIBA1987427,tsuboi1988magnetization,hori1989magnetization}, which lie beyond the explanatory power of the spin-1 framework. While this behavior points to the need for a fundamentally different low-energy description, its microscopic origin has remained unresolved for nearly four decades. Moreover, as demonstrated in the high-pressure measurements presented later in this work, the magnetization plateaus exhibit even greater complexity, further enriching the phenomenology of the high-field regime.

To address this, we introduce a perspective in which the magnetization process of CsFeCl\(_3\) is naturally divided into two regimes, each described by a distinct effective spin model. The low- to intermediate-field regime is captured by an effective spin-1 Hamiltonian, denoted \( \mathcal{H}_{\rm eff}^{\rm spin\text{-}1} \). In contrast, the high-field regime near the metamagnetic response, featuring quantized magnetization steps and plateaus, is described by a projected spin-1/2 model, \( \mathcal{H}_{\rm eff}^{\rm spin\text{-}1/2} \), constructed from {the two lowest-energy crystal-field levels that are selected by the Zeeman effect at high fields, namely} \( m_{J2}^z = +2 \) and \( m_{J1}^z = +1 \). In this projected model, the conventional symmetry \( \mathcal{H}(\{\bm{S}_i\}, \bm{H}) = \mathcal{H}(\{-\bm{S}_i\}, -\bm{H}) \), common in standard spin Hamiltonians, no longer holds. This breakdown occurs because the two effective spin states are not related by time reversal, as they belong to different crystal-field multiplets. Consequently, the emergent spin model intrinsically breaks TRS and permits odd-body interaction terms, such as three-body couplings on triangular plaquettes, that are typically forbidden in conventional systems. These features give rise to asymmetric magnetization plateaus under field reversal, as illustrated schematically in Fig.~\ref{sketch}(b), and provide a natural resolution to the long-standing puzzle of anomalous high-field metamagnetism in CsFeCl$_3$~\cite{CHIBA1987427,tsuboi1988magnetization}. This mechanism may therefore also be applicable to other hexagonal $ABX_3$-type magnets, such as RbFeCl$_3$~\cite{Amaya1988-pz,Stoppel2021-vw}, which exhibits similar metamagnetic behavior in high magnetic fields.

Below, we establish this scenario through a combination of experimental and theoretical approaches. We performed high-field magnetization and electron spin resonance (ESR) measurements over a wide frequency-field range as well as magnetic susceptibility experiments under high pressure. On the theoretical side, we construct effective spin models for both the low- and high-field regimes, \( \mathcal{H}_{\rm eff}^{\rm spin\text{-}1} \) and \( \mathcal{H}_{\rm eff}^{\rm spin\text{-}1/2} \), using model parameters determined by fits to experimental data. While our analysis provides a unified understanding of the magnetization process in CsFeCl\(_3\), its implications are more general. This framework naturally extends to other quantum magnets with multiple low-lying crystal-field states, where a similar mechanism can lead to emergent TRS breaking in the high-field low-energy description. Our findings, therefore, open a broad and largely unexplored landscape of effective spin models enriched by unconventional odd-body interaction terms, offering new directions for both theoretical and experimental studies of quantum magnetism.

The format of this paper is as follows. In Sec. II, the sample preparation of single crystals of CsFeCl$_3$, experimental methods and conditions are described. In Sec. III, we show the experimental results of the magnetization, ESR, and high-pressure magnetic susceptibility of CsFeCl$_3$. In Sec. IV, we discuss the pressure dependence of the magnetic transition fields and analyze the mechanism of the metamagnetic transition through comparison with effective spin models. Finally, we summarize and conclude our study.

\section{Experimental}

Single crystals of CsFeCl$_3$ were grown in an evacuated quartz crucible by using the Bridgman method~\cite{kurita2016magnetic}. The obtained single crystals were confirmed to be CsFeCl$_3$ by X-ray diffraction. Magnetic fields were applied along the $c$-axis of CsFeCl$_3$ single crystals within an accuracy of a few degrees. Magnetization ($M$) of this sample at ambient pressure was measured by an induction method using a pick-up coil at 1.4 K in pulsed magnetic fields of up to 51 T generated by a non-destructive pulse magnet and a capacitor bank installed at Center for Advanced High Magnetic Field Science (AHMF), Graduate School of Science, The University of Osaka. ESR measurements at 4.2 K in pulsed high magnetic fields were employed with a far-infrared laser (Edinburgh Instruments, UK)(frequency range: 326-1839 GHz) at AHMF, and with VDI microwave-chain radiation sources (product of Virginia Diodes, Inc., USA)(60-120 GHz) at Dresden High Magnetic Field Laboratory (HLD), Helmholtz-Zentrum Dresden-Rossendorf (HZDR).

To investigate the magnetization response under high pressure, we utilized a proximity detector oscillator (PDO) system combined with a piston-cylinder pressure cell (PCC) which allows us to use the pulse magnet at AHMF~\cite{nihongi2023piston}. The PDO system is a self-resonating inductance-capacitance (LC) circuit based on a proximity detector chip~\cite{Altarawneh}. When a magnetic field is applied to a magnetic insulator inside an inductor, the magnetic susceptibility ($\Delta M/ \Delta H$) of the sample induces a change in the inductance, leading to a change in the resonance frequency ($\Delta f$)~\cite{Ghannadzadeh}. In this setup, the sample placed inside the inductor was inserted into the PCC filled with the pressure medium Daphne 7373. The resonance frequency $f$ in the circuit was approximately 37 MHz at zero field. The pressure was calibrated with the superconducting transition temperature of tin which was placed in the PCC. The details of this method are described in Ref.~\cite{nihongi2023piston}.

\section{Results}
\subsection{Magnetization at ambient pressure}

\begin{figure}[htb]
\begin{center}
\includegraphics[keepaspectratio, scale=0.85]{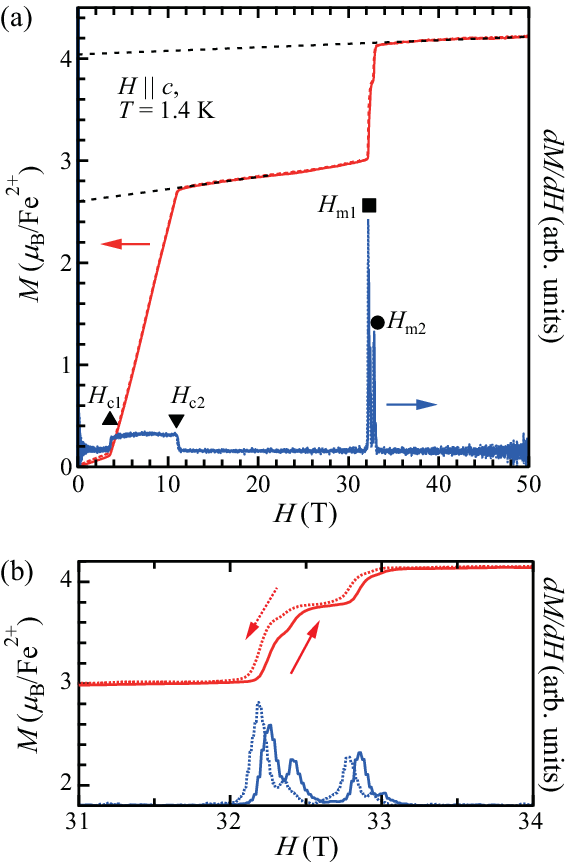}
\caption{(a) Magnetization ($M$) and its field derivative ($dM/dH$) of CsFeCl$_3$ at 1.4 K for $H \parallel c$ at ambient pressure. Broken lines are the extrapolations of the magnetization slope above $H_{\rm c2}$ and $H_{\rm m2}$ toward zero field. (b) Enlarged view of $M$ and $dM/dH$ around $H_{\rm m1}$ and $H_{\rm m2}$. The arrows indicate the field-ascending and descending processes.}\label{HFMAG}
\end{center}
\end{figure}

\begin{figure}[htb]
\includegraphics[keepaspectratio, scale=0.85]{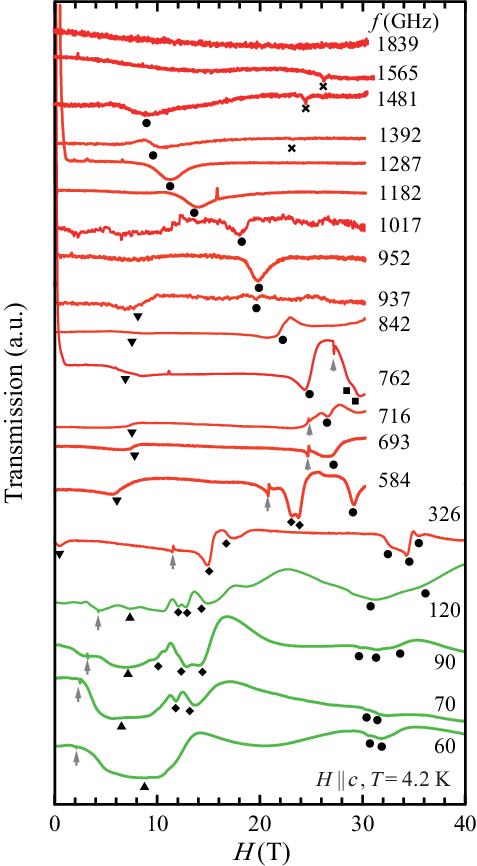}
\caption{ESR absorption spectra of CsFeCl$_3$ at 4.2 K for $H \parallel c$ taken between 326 and 1839 GHz (red) at AHMF and between 60 and 120 GHz (green) at HLD. As the transmission of the spectra at 1017, 1481, 1565, and 1839 GHz is weak, their spectra are plotted by a factor of 10 times larger than the other spectra. The resonance fields are indicated by upward and downward triangles, diamonds, circles, squares, and crosses. Gray arrows point to the signal of the ESR standard marker DPPH (2,2-diphenyl-1-picrylhydrazyl).}
\label{ESR}
\end{figure}

\begin{figure}[htb]
\includegraphics[keepaspectratio, scale=0.9]{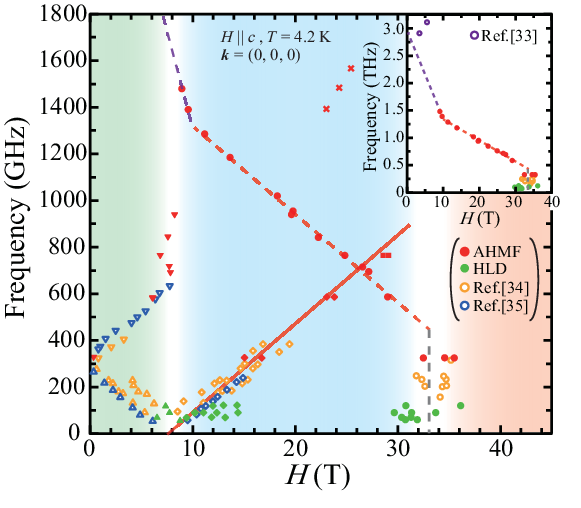}
\caption{Frequency vs resonance-field plot obtained by ESR measurements. Red and green symbols are the resonance fields taken from Fig. \ref{ESR}. Open colored symbols are resonance fields taken from Refs.~\onlinecite{motokawa1989submillimeter, mitsudo2003high, okubo2023high}. The red solid line indicates the linear fit of the solid diamond symbols corresponding to the transitions between the ground $m_{J1}^z$ = $+1$ and the excited $m_{J1}^z$ = 0 states(see text). The red and purple broken lines are linear fits to the red circles at 326-1287 GHz and 1392-1481 GHz, respectively. The field at the gray broken line is approximately 33 T, which is close to the metamagnetic transition field. The inset indicates the resonance fields related to the transition between the ground state and $m_{J2}^z$ = +2 state. The ground state in magnetic fields is considered to change from  $m_{J1}^z$ = 0 to $m_{J1}^z$ = $+1$, and then to $m_{J2}^z$ = $+2$ as indicated by green, blue, and red hatching, respectively. These resonance fields were obtained at the ESR wave vector $\mbox{\boldmath $k$}$ = (0, 0, 0).}
\label{ESR_diagram}
\end{figure}

Figure~\ref{HFMAG}~(a) shows $M$ and its field derivative ($dM/dH$) of CsFeCl$_3$ at 1.4 K for $H \parallel c$ in pulsed magnetic fields at ambient pressure. $M$ increases gradually up to $H_{\rm c1}$ = 3.7 T, which corresponds to a quantum phase transition from the $m_{J1}^z$ = 0 to $m_{J1}^z$ = +1 states, and then increased rapidly between $H_{\rm c1}$ and $H_{\rm c2}$ = 10.7 T. Above $H_{\rm c2}$, $M$ further grows with a small slope with increasing magnetic field, which is attributed to Van Vleck paramagnetism, and then a metamagnetic transition with a step appears at $H_{\rm m1}$ = 32.2 T and $H_{\rm m2}$ = 32.8 T. These features are in good agreement with those in the previous studies~\cite{CHIBA1987427, tsuboi1988magnetization}. As visible in the enlarged view of $M$ and $dM/dH$ in Fig.~\ref{HFMAG}(b), the metamagnetic transition exhibits a small hysteresis, indicating a first-order transition, and includes a narrow step-like magnetization plateau. The total metamagnetic-step size at $H_{\rm m1}$ and $H_{\rm m2}$ is approximately 1.1 $\mu_{\rm B}$/Fe$^{2+}$ in both field-ascending and descending processes. We observe another peak in $dM/dH$ at 32.4 T during the field-ascending process that is smaller in the field-descending process. Assuming that $H_{\rm c2}$ corresponds to the saturation field of the $J$ = 1 state with an effective spin $s$ = 1, the intercept of the magnetization slope above $H_{\rm c2}$ with $H$ = 0 is $M$ = $g\mu_{\rm B}s$ = 2.61 $\mu_{\rm B}$/Fe$^{2+}$. The $g$ value is consistent with $g$ = 2.6 obtained from the previous ESR measurements~\cite{motokawa1989submillimeter, mitsudo2003high, okubo2023high}. $M$ extrapolated above $H_{\rm m2}$ intersects at $M$ = 4.05 $\mu_{\rm B}$/Fe$^{2+}$. Assuming that $M$ above $H_{\rm m2}$ corresponds to the saturation magnetization of the $J$ = 2 state, the $g$ value is estimated to be 2.03.
\begin{figure*}[htb]
\begin{center}
\includegraphics[keepaspectratio, scale=0.9]{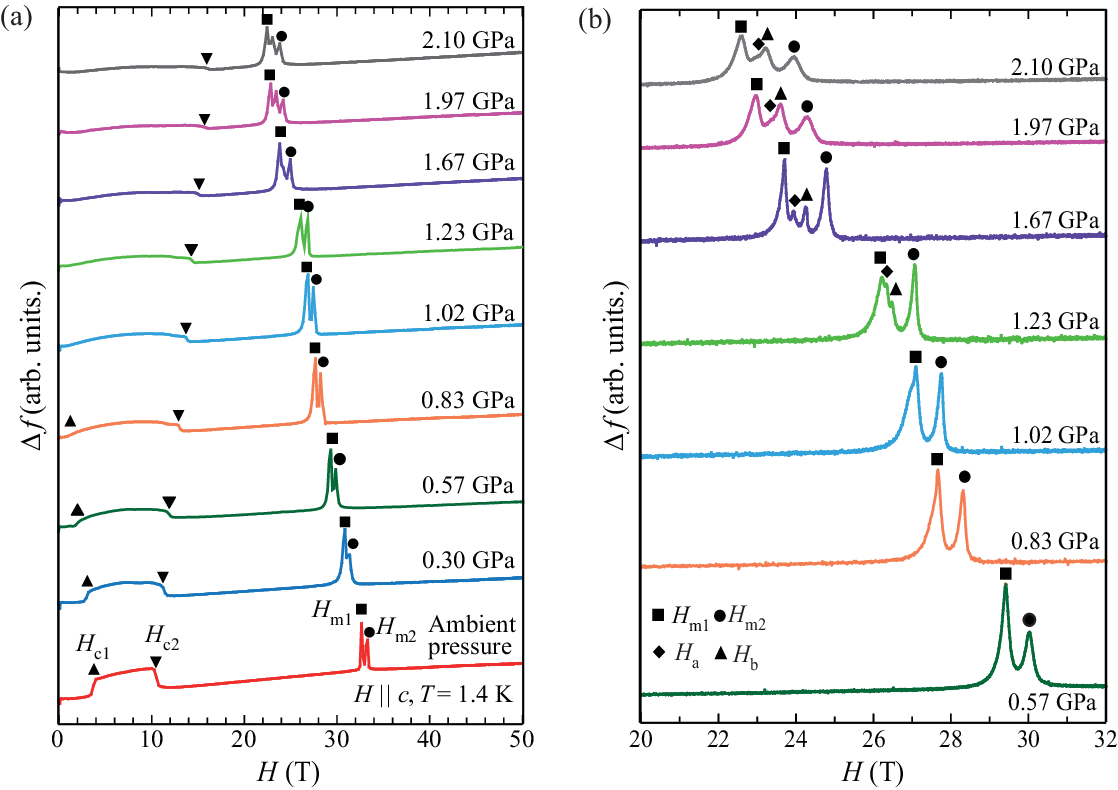}
\caption{(a) Change in the resonance frequency ($\Delta f$) of CsFeCl$_3$ at 1.4 K for $H \parallel c$ under various pressures. (b) Enlarged view of $\Delta f$ around the metamagnetic transition fields above 0.57 GPa.}\label{PDO_1}
\end{center}
\end{figure*}
\begin{figure*}[t]
\begin{center}
\includegraphics[keepaspectratio, scale=0.88]{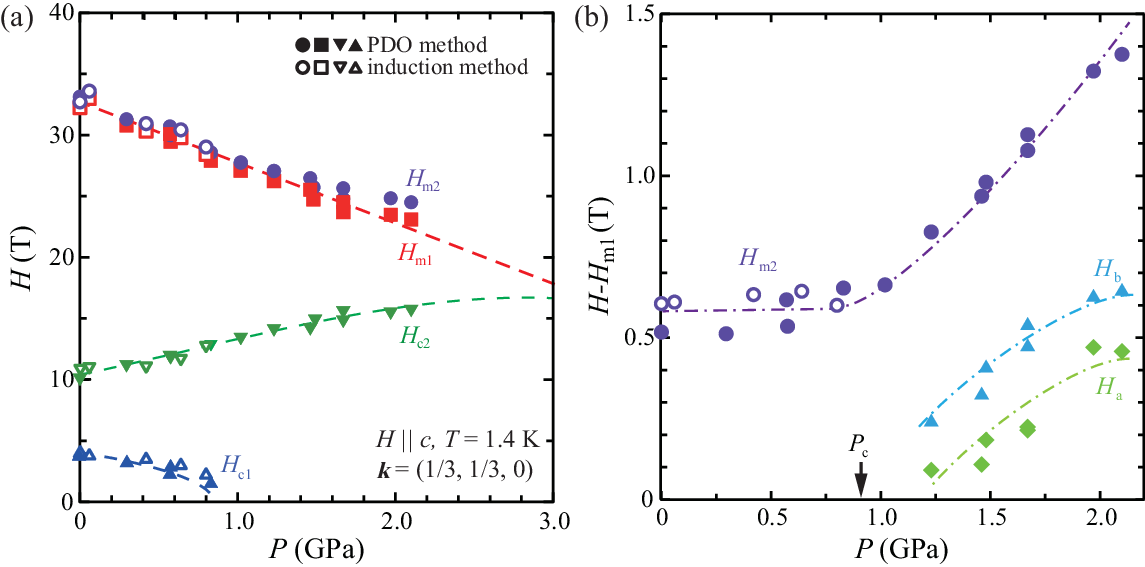}
\caption{(a) Magnetic field vs. pressure phase diagram of CsFeCl$_3$ at 1.4 K for $H \parallel c$. Solid and open symbols are experimental data obtained using the PDO and the induction method~\cite{nihongi2023high}, respectively. The blue and green broken curves are the pressure dependences of $H_{\rm c1}$ and $H_{\rm c2}$ determined from Eqs.~(\ref{H_c1_c2_mean_theory}) and~(\ref{H_c1_c2_mean_theory2}). The red broken line is a fit to a metamagnetic transition fields. (b) Difference between $H_{\rm m1}$ and other metamagnetic transition fields ($H_{\rm m2}$, $H_{\rm a}$, $H_{\rm b}$) as a function of pressure. The dot-dashed lines are the guides to the eyes.\label{phase_diagram}}
\end{center}
\end{figure*}

\subsection{Electron spin resonance at ambient pressure}

Figure~\ref{ESR} shows ESR absorption spectra of CsFeCl$_3$ at 4.2 K for $H \parallel c$ at 60-1839 GHz. The resonance fields presumably corresponding to the same transitions are denoted by the same symbols. The tiny sharp resonance absorptions indicated by gray arrows correspond to the ESR absorption spectra of DPPH (2,2-diphenyl-1-picrylhydrazyl), which is an ESR standard marker with $g$ = 2.0036. The resonance field indicated by the downward triangle shifts from almost zero field at 326 GHz to 7.9 T at 693 GHz, and then becomes nearly constant above 693 GHz. The resonance field indicated by the upward triangle below 120 GHz seems to shift slightly to a lower field with increasing frequency. On the other hand, the resonance field indicated by the diamonds, which appears above 70 GHz, shifts to a higher magnetic fields with increasing frequency. The absorption spectra indicated by circles at around 32 T below 326 GHz represent a few resonance absorption fields. These resonance fields are scattered around the metamagnetic transition field.  Above 584 GHz, the resonance field indicated by the circle shifts to lower magnetic fields with increasing frequency. Above 1565 GHz, the ESR line is not resolvable anymore because of strong absorption broadening. A small ESR absorption signal above 1392 GHz indicated by crosses shifts to higher magnetic fields with increasing frequency up to 1565 GHz and disappears at 1839 GHz.

A frequency vs. resonance-field plot is shown in Fig.~\ref{ESR_diagram}. In this plot, we included the resonance fields reported in previous ESR studies~\cite{motokawa1989submillimeter,mitsudo2003high, okubo2023high}, indicated by open colored symbols. The up- and downward triangles correspond to the transitions between the ground $m_{J1}^z$ = 0 and the excited $m_{J1}^z$ = $\mp 1$ states. The diamonds correspond to the transitions between the new field-induced ground state $m_{J1}^z$ = +1 and the excited $m_{J1}^z$ = 0 state. Based on the frequency dependence of the upward triangles and diamonds, the transition field between $m_{J1}^z$ = 0 and $m_{J1}^z$ = +1 states is estimated to be around 8 T, which is close to the frequency-independent (FI) transition field indicated by the downward triangles above 693 GHz. This FI transition must be related to a change in the dielectric constant at the phase transition.
The resonances depicted by crosses above 1392 GHz can be described as 2$g$ = 5.17. Thus, the resonances must correspond to the transition between the ground $m_{J1}^z$ = +1 and the excited $m_{J1}^z$ = -1 states.

The red circles below approximately 30 T are located at higher frequencies than the diamonds. Referring to the energy level of CsFeCl$_3$ in Fig.~\ref{f1}(b), these circles are considered to be transitions between the ground state and the $m_{J2}^{z}$ = +2 states. A linear fit between 584 and 1287 GHz (red broken line in Fig.~\ref{ESR_diagram}) yields a slope of $g$ = 2.70. The frequency dependence of the resonances marked by circles between 1392 and 1481 GHz is different from that below 1392 GHz. As shown in the inset of Fig.~\ref{ESR_diagram}, the straight line drawn through the two circles at 1392 and 1481 GHz intersects at approximately 2.8 THz at zero field. This straight line is located near the resonance fields reported previously~\cite{motokawa1989submillimeter} and could correspond to the transition between the $m_{J1}^{z}$ = 0 and the $m_{J2}^{z}$ = +2 states. The intersecting magnetic field between red and purple broken lines around 1300 GHz coincides with the transition field ($\sim$ 8 T) between the $m_{J1}^{z}$ = 0 and the $m_{J1}^{z}$ = +1 states. Therefore, the purple and red broken lines are suggested to be the ESR modes corresponding to the transitions between the ground state ($m_{J1}^z$ = 0 below the transition field and $m_{J1}^{z}$ = +1 above the transition field) and the excited $m_{J2}^{z}$ = +2 states at wave vector $\mbox{\boldmath $k$}$ = ($h$, $k$, $l$) = (0, 0, 0). (The definition will be given later.) This transition field is different from $H_{\rm c1}$, which should be the transition at $\mbox{\boldmath $k$}$ = (1/3, 1/3, 0) $\equiv$~$\mbox{\boldmath $Q$}$.

The scattered circles around 33 T, which is close to the metamagnetic transition field, indicate that the ESR mode drops rapidly from approximately 400 GHz. This suggests that the metamagnetic transition is caused by the level crossing between the $m_{J1}^{z}$ = +1 and the $m_{J2}^{z}$ = +2 states at $\mbox{\boldmath $k$}$ = $\mbox{\boldmath $Q$}$. The green solid symbols measured at HLD are higher for the resonance fields around 12 T (lower for those around 32 T) than those open symbols measured at other places. This probably is caused by a small misalignment of the sample against the external field.
 
\subsection{Magnetic susceptibility using a PDO  under pressure}

Figure~\ref{PDO_1}~(a) shows the $\Delta f$-$H$ curves of CsFeCl$_3$ at 1.4 K for $H \parallel c$ at various pressures. At ambient pressure, we observe shoulders at $H_{\rm c1}$ and $H_{\rm c2}$ and two peaks at $H_{\rm m1}$ and $H_{\rm m2}$. The additional peak between $H_{\rm m1}$ and $H_{\rm m2}$ observed in $dM/dH$, did not appear in the field-ascending process. $H_{\rm c1}$ shifts to lower magnetic fields with increasing pressure and disappeared above 1.02 GPa, which agrees with the critical pressure $P_{\rm c} \approx$ 0.9 GPa in Ref.~\onlinecite{kurita2016magnetic}. In contrast, $H_{\rm c2}$ shifts to higher magnetic field, and $H_{\rm m1}$ and $H_{\rm m2}$ decrease with increasing pressure up to highest pressure. These results are in good agreement with our magnetization measurements using the induction method under pressure $\sim$0.80 GPa~\cite{nihongi2023high}. Figure~\ref{PDO_1}~(b) displays an enlarged view of $\Delta f$ data around the metamagnetic transitions above 0.57 GPa. Above 1.23 GPa, two additional peaks in $\Delta f$ appear between $H_{\rm m1}$ and $H_{\rm m2}$. We label the low and high transition fields as $H_{\rm a}$ and $H_{\rm b}$, respectively. The peaks at $H_{\rm a}$ and $H_{\rm b}$ seem to separate from the anomaly at $H_{\rm m1}$. At 1.67 GPa, the peaks at $H_{\rm a}$ and $H_{\rm b}$ are clearly separated. At 1.97 GPa and 2.10 GPa, the peaks merge.

Figure~\ref{phase_diagram}~(a) shows the magnetic-field vs. pressure phase diagram of CsFeCl$_3$ at 1.4 K for $H \parallel c$ obtained by our magnetic susceptibility and magnetization measurements~\cite{nihongi2023high} in pulsed high magnetic fields under high pressures. With increasing pressure, $H_{\rm c1}$ decreases as approaching $P_{\rm c}$, $H_{\rm m1}$ and $H_{\rm m2}$ decrease as well continuously, whereas $H_{\rm c2}$ increases monotonically. Accordingly, $H_{\rm c2}$ and $H_{\rm m1}$ might intersect above 3 GPa. The details of the fittings will be described in the next section.

To visualize the relationship between the lowest metamagnetic transition field $H_{\rm m1}$ and the other transition fields ($H_{\rm m2}$, $H_{\rm a}$, $H_{\rm b}$), the difference between $H_{\rm m1}$ and the other fields vs. pressure is shown in Fig.~\ref{phase_diagram}~(b). The difference between $H_{\rm m1}$ and $H_{\rm m2}$ is almost constant below 0.83 GPa and increases linearly with increasing pressure above 1.02 GPa. These features suggest that the metamagnetic transition fields relate to $P_{\rm c}$. Moreover, $H_{\rm a}$ and $H_{\rm b}$ move away from $H_{\rm m1}$ with increasing pressure. The widening of the difference between $H_{\rm m1}$ and $H_{\rm m2}$ corresponds to a broadening of the metamagnetic transition, which might indicate the enhancement of exchange interactions with increasing pressure.

\subsection{Pressure effect of the metamagnetic transition}

We analyze the successive metamagnetic transitions and their pressure effects. Figure \ref{Integral_f}~(a) shows $M$ normalized by the magnitude of the metamagnetic transition and $dM/dH$ of CsFeCl$_3$ at 1.4 K for $H \parallel c$ at ambient pressure. The step-like metamagnetic transitions during the field-ascending process exhibit magnetization plateaus around 1/3 and 2/3 of the normalized saturation magnetization. The plateau width around 1/3 is much smaller than that around 2/3. Figures~\ref{Integral_f}~(b)-\ref{Integral_f}~(f) show $\Delta f_{\rm sub}$-$H$ curves (lower panels) and their integrated values (upper panels) as a function of magnetic field around the metamagnetic transition under various pressures. 

To compare the magnetization change at metamagnetic transitions, $\Delta f_{\rm sub}$ was obtained by adjusting the value of $\Delta f$ before and after the metamagnetic transition field to zero. The field integrated $|\Delta f_{\rm sub}|$ corresponds to $M$ and its integrated value was normalized by the saturation value. At 0.83 GPa and 1.02 GPa, the integrated $|\Delta f_{\rm sub}|$ shows a plateau around 2/3 but no step around 1/3. Above 1.23 GPa, the two stepwise plateaus change into the four step-like features corresponding to the change from two to four peaks in $|\Delta f_{\rm sub}|$. Accordingly, we observe approximate 1/3, 1/2 plateaus,~together with a plateau appearing around $2/3$ or $3/4$ of the saturation magnetization. At 2.10 GPa, all magnetization plateaus are obscured due to the blunt peaks in $|\Delta f_{\rm sub}|$.  

Next, we analyze the pressure dependence of the metamagnetic transition fields ($H_{\rm m1}$ and $H_{\rm m2}$). The outcomes of the ESR measurements evidences that the metamagnetic transition is caused by the level-crossing between $m_{J1}^{z}$ = +1 and $m_{J2}^{z}$ = +2 states. Considering the energy levels of the Fe$^{2+}$ ion in Fig.~\ref{f1}~(b), the decrease in the metamagnetic transition fields with increasing pressure can be explained by a pressure effect on the single-ion anisotropy ($D$) in the lowest $J$ = 1 triplet states and the single-ion anisotropy ($D_{\rm e}$) in the excited $J$ = 2 quintet states. $D$ and 4$D_{\rm e}$ correspond to the energy difference between $m_{J1}^{z}$ = $\mp 1$ and $m_{J1}^{z}$ = 0, and $m_{J2}^{z}$ = 0 and $m_{J2}^{z}$ = $\mp 2$, respectively. These $D$ and $D_{\rm e}$ values may depend on $\delta/\lambda\rq{}$ where $\lambda\rq{}$(= $k\lambda$) is an effective spin-orbit coupling constant~\cite{inomata1967theory}. Assuming that the Zeeman effect of each energy level remains unchanged with pressure, the decrease in the metamagnetic transition fields indicates the decrease of the energy difference between $m_{J1}^{z}$ = $\mp 1$ and $m_{J2}^{z}$ = $\mp 2$ with increasing $\delta/\lambda\rq{}$. In our previous study~\cite{nihongi2023high}, our data suggest a weak pressure dependence of $\lambda\rq{}$. Therefore, the pressure dependence of $H_{\rm m1}$ and $H_{\rm m2}$ indicate that the trigonal field ($\delta$) of the FeCl$_6$ octahedra becomes enhanced with increasing pressure, leading to an increase of the single-ion anisotropies $D$ and $D_{\rm e}$. The increase in $D$ with the pressure is consistent with the result in Ref.~\cite{hayashida2019novel}. We note that the metamagnetic transition depends not only on the single-ion anisotropy but also on the exchange interactions, as shown below, similar to $H_{\rm c1}$ and $H_{\rm c2}$.

\section{Discussion}
\subsection{Effective model of the low- to intermediate-field regime}

As described in the introduction, at low temperatures far below $|\lambda|/k_{\rm B} \simeq 150 $ K, the magnetic properties are characterized by the $J$ = 1 state. Since the ground state is a singlet with $m_{J1}^z$ = 0, CsFeCl$_3$ shows a non-magnetic disordered state. The spin Hamiltonian is given by

\begin{eqnarray}
\mathcal{H}_{\rm eff}^{\rm spin\text{-}1}  &= &~\sum_i D(s_i^z)^2 -J_{0} \sum_{<i,j>}^{\rm chain}\left[(s_i^xs_j^x+s_i^ys_j^y)+\xi (s_i^zs_j^z) \right] \nonumber \\
&+&J_1\sum_{<i,j>}^{\rm plane}\left[(s_i^xs_j^x+s_i^ys_j^y)+\xi(s_i^zs_j^z) \right] - g\mu_{\rm B}\sum_i s_i^z H,
\label{matsumoto_model}
\end{eqnarray}
where $s$ is the effective spin one. The magnitude of the single-ion anisotropy is positive ($D > 0$) and corresponds to the energy gap between $m_{J1}^{z}$ = 0 and $m_{J1}^{z}$ = $\mp 1$ as shown in Fig.~\ref{f1}~(b). The Hamiltonian includes two exchange terms: the ferromagnetic exchange interaction $J_0$ in the Fe$^{2+}$ chain along the $c$-axis and the antiferromagnetic exchange interaction $J_1$ in the triangular-lattice layer parallel to the $ab$-plane~\cite{yoshizawa1980neutron}. $\xi$ denotes the exchange anisotropy. The last term is the Zeeman energy with $g$ value at $J$ = 1 state. Note that due to the negative sign of the Zeeman energy term, the energy state of spins with a positive sign decreases with increasing magnetic field.

\begin{figure*}[t]
\begin{center}
\includegraphics[keepaspectratio, scale=0.5]{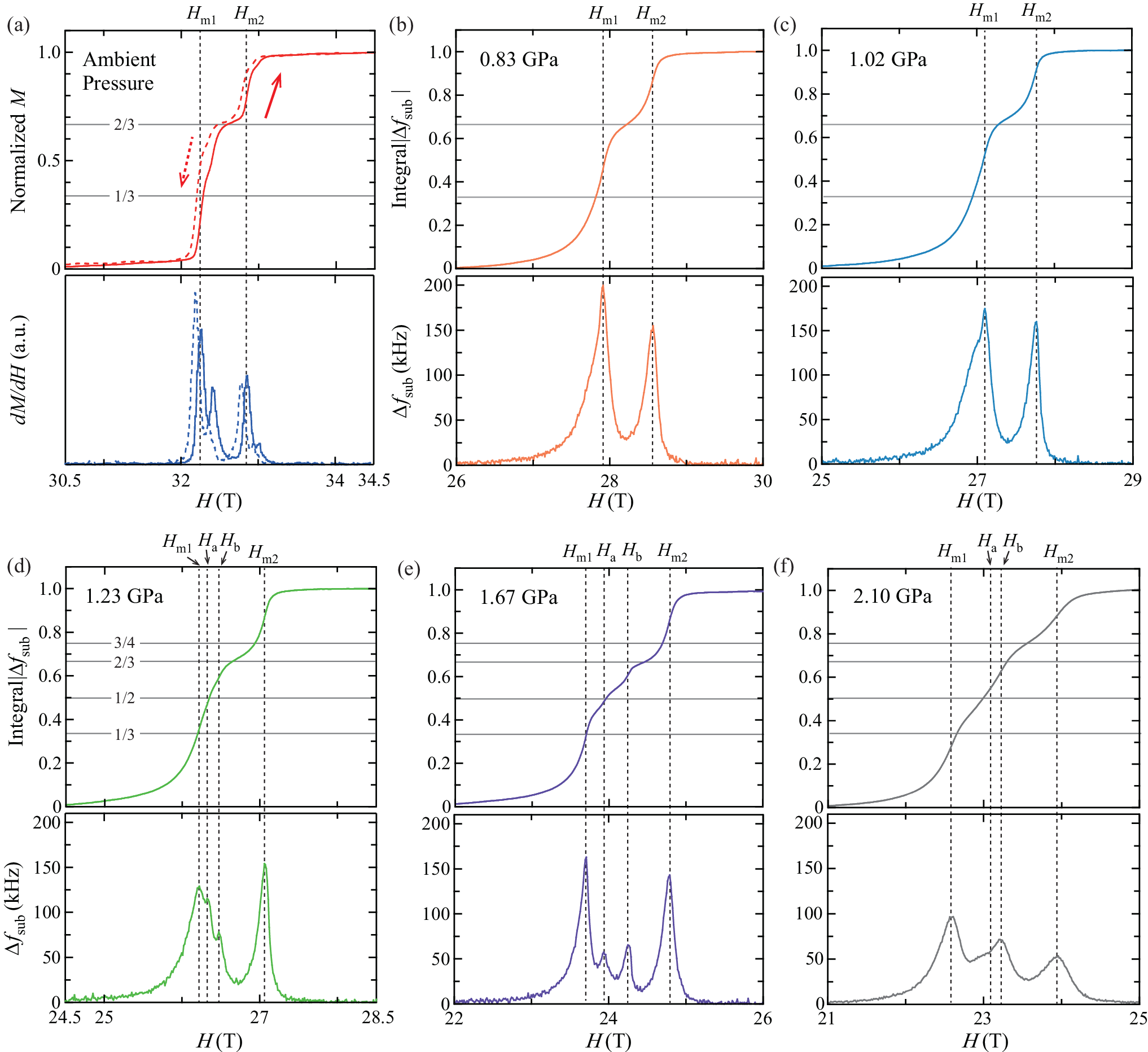}
\caption{(a) $M$ normalized by the magnitude of the metamagnetic transition (upper panel) and $dM/dH$ (lower panel) at ambient pressure measured with the induction method. (b)-(f) Field integrated $|\Delta f_{\rm sub}|$ (upper panels) and $\Delta f_{\rm sub}$ (lower panels) at the pressures indicated in each panel.}\label{Integral_f}
\end{center}
\end{figure*}

First, the pressure dependence of $H_{\rm c1}$ and $H_{\rm c2}$ is analyzed. $H_{\rm c1}$ and $H_{\rm c2}$ are explained within the framework of the $J$ = 1 state, which is replaced as a fictitious spin $s$ =1 and consists of the singlet ground state with $s_{z}$ = 0 and the excited doublet state with $s_{z}$ = $\pm 1$, separated by $D$. The excited state possesses a finite energy band owing to the exchange interaction. The lower doublet state with a finite bandwidth moves down to cross the singlet ground state between $H_{\rm c1}$ and $H_{\rm c2}$. Below $H_{\rm c1}$, the dispersion relation of the excited state is expressed by using the Hamiltonian as follows:~\cite{matsumoto2020analysis}
\begin{equation}
E_{H<H_{\rm c1}}^{\pm} = \sqrt{D[D+2(-2J_0{\rm cos}\pi l+2 J_1\gamma(\mbox{\boldmath $k$}))]}\pm g\mu_{\rm B}H
\label{matsumoto_model2}
\end{equation}
\noindent
and
\begin{align}
\gamma(\mbox{\boldmath $k$}) & =~{\rm cos} 2\pi h + {\rm cos} 2\pi k + {\rm cos}2\pi (h + k), \nonumber \\
\mbox{\boldmath $k$} & =~h\mbox{\boldmath $a$}^* + k\mbox{\boldmath $b$}^* + h\mbox{\boldmath $c$}^*.
\end{align}
\noindent
Here, $\pm$ of $E^\pm_{H<Hc1}$ represents the higher and lower excited states, respectively. The second term $\pm$ $g \mu_{\rm B}H$ is the Zeeman splitting of the excited doublet state with $s_{z}$ = $\pm$1, and $\mbox{\boldmath $a$}^*$, $\mbox{\boldmath $b$}^*$, and $\mbox{\boldmath $c$}^*$ are reciprocal lattice vectors. At wave vector $\mbox{\boldmath $k$}$ = ($h$, $k$, $l$) = (1/3, 1/3, 0)~$\equiv$~$\mbox{\boldmath $Q$}$, $\gamma$($\mbox{\boldmath $k$}$) takes the minimum value $\gamma$ ($\mbox{\boldmath $Q$}$) = -3/2. Then, the excitation gap is given by
\begin{equation}
E_{H<H_{\rm c1}}^{\rm gap} = E_{H<H_{\rm c1}}(\mbox{\boldmath $Q$}) = \sqrt{D[D-2(2J_0+3J_1)]}- g\mu_{\rm B}H.
\label{matsumoto_model2}
\end{equation}
\noindent
Accordingly, $H_{\rm c1}$, where the $E_{H<H_{\rm c1}}^{\rm gap}$ becomes zero, is written as,
\begin{eqnarray}
H_{\rm c1} = \sqrt{D^2-2D(2J_0+3J_1)}/g\mu_{\rm B}.
\label{H_c1_c2_mean_theory}
\end{eqnarray}
\noindent

By substituting the pressure dependence of the exchange constants ($J_0$, $J_1$) and the single-ion anisotropy constant ($D$) obtained from the inelastic neutron scattering (INS) experiments ($J_0 = 0.5 + 0.14P$ (meV), $J_1 = 0.0312 - 0.0015P$ (meV), and $D = 2.345 + 0.365P$ (meV),
where $P$ (GPa) is the pressure value~\cite{hayashida2019novel}) into Eq.~(\ref{H_c1_c2_mean_theory}), the pressure dependence of $H_{\rm c1}$ is given as the blue broken line in Fig.~\ref{phase_diagram}. The calculated data of $H_{\rm c1}$ is in good agreement with the experimental data. 

Above $H_{\rm c2}$, the $s$ = 1 spins align in the direction of the magnetic field. Accordingly, the ground state, the lowest excited state, and the second lowest excited state are the $s_{z}$ = -1, 0, and +1 states, respectively. The dispersion relation of the excited $s_{z}$ = 0 state is written as,
\begin{equation}
E_{H>H_{\rm c2}}(\mbox{\boldmath $k$}) = -D+2J_0^z(\xi-{\rm cos}k_z)+2 J_1\gamma( \mbox{\boldmath $k$})-6\xi J_1+ g\mu_{\rm B}H.
\label{matsumoto_model4}
\end{equation}
\noindent
The excitation minimum is located at $\mbox{\boldmath $k$}$ = $\mbox{\boldmath $Q$}$, and the excitation gap is given by,
\begin{equation}
E_{H>H_{\rm c2}}^{\rm gap} = E_{H>H_{\rm c2}}(\mbox{\boldmath $Q$}) = g\mu_{\rm B}H - D + 2 J_0 (\xi-1) -3J_1 (1 + 2 \xi).
\label{matsumoto_model2}
\end{equation}
\noindent
Since ESR detects the signals corresponding to the transitions between the ground state and the excited state at $\mbox{\boldmath $k$}$ = 0, the excitation gap between the $s_z$ = -1 and the $s_z$ = 0 states is written as,
\begin{equation}
E_{H>H_{\rm c2}}^{\rm ESR} = E_{H>H_{\rm c2}}(0) = g\mu_{\rm B}H - D + 2 J_0 (\xi-1) +6J_1 (1 - \xi).
\label{matsumoto_model3}
\end{equation}
\noindent
The diamonds in Fig.~\ref{ESR_diagram} correspond to the transitions between the $s_{z}$ = -1 ($m_{J1}^{z}$ = +1) and the $s_{z}$ = 0 ($m_{J1}^{z}$ = 0) states, and they are described by Eq. (\ref{matsumoto_model3}). The Parameters are based on the exchange interaction and anisotropy described above at $P$ = 0 ~\cite{hayashida2019novel} and $g$ = 2.70 obtained by our ESR results. The exchange anisotropy at ambient pressure is $\xi$ = 2.40, which is determined by the fitting~{Eq.~\ref{matsumoto_model3} to the red diamonds (red solid line) in Fig.~ \ref{ESR_diagram}.

$H_{\rm c2}$ corresponds to the magnetic field at $E_{H>H_{\rm c2}}^{\rm gap}$ = 0, and it is given by the following equation using $\xi$ in addition to $D$, $J_0$, and $J_1$,
\begin{equation}
H_{\rm c2} = \{D-2J_0(\xi-1)+3J_1(1+2\xi)\}/g\mu_{\rm B}.
\label{H_c1_c2_mean_theory2}
\end{equation}
\noindent
The green broken line in Fig.~\ref{phase_diagram}(a) expresses the pressure dependence of $H_{\rm c2}$ given by Eq.~(\ref{H_c1_c2_mean_theory2}), in which we used the pressure dependence of $D$, $J_0$, and $J_1$ determined by INS experiments~\cite{hayashida2019novel} and employed $g$ = 2.61 obtained from the magnetization data. In this fitting, the pressure dependence of $\xi$ is expressed as,
\begin{equation}
\xi (P) = 2.29 -0.52P +0.11 P^2.
\label{pre_dep_xi}
\end{equation}
\noindent $\xi$ at ambient pressure ($P$ = 0) is in agreement with that obtained by our ESR measurement. The pressure dependence of $\xi$ indicates that the exchange anisotropy becomes small with increasing pressure up to the highest pressure measured.

As described above, the pressure dependence of the magnetization process in the low- to intermediate-field region can be consistently described by the spin-1 $XXZ$ + $D$ model, $\mathcal{H}_{\rm eff}^{\rm spin\text{-}1}$, given in Eq.~\eqref{matsumoto_model}. For pressures below the critical value, $P_{\mathrm{c}} \approx 0.9~\mathrm{GPa}$, the system evolves from a singlet non-magnetic ground state to a magnetically ordered phase at $H_{\mathrm{c1}}$ via magnon BEC~\cite{PhysRevLett.84.5868,PhysRevB.71.224426,Giamarchi2008,RevModPhys.86.563}, exhibiting an umbrella-type spin configuration, and reaches an apparent ``saturation'' at $H_{\mathrm{c2}}$ within the spin-1 framework. Above $P_{\mathrm{c}}$, the relative reduction of the single-ion anisotropy $D$ with respect to exchange interactions closes the excitation gap and eliminates the non-magnetic phase at low fields. The pressure dependence of the model parameters reproducing this behavior were determined within a linear approximation, using previous INS data~\cite{hayashida2019novel} together with the present $H_{\mathrm{c2}}$ measurements. Further refinement of these parameters should be possible by combining the present analysis with more detailed experimental studies, including high-pressure magnetization measurements.


\color{black}

\subsection{Effective model of the metamagnetic-transition-field regime}

While the low- to intermediate-field regime has been explained within the model (\ref{matsumoto_model}) with a fictitious spin 1, corresponding to the three low-lying eigenstates of the $J=1$ manifold, the metamagnetic-transition-field regime should be modeled by a fictitious (pseudo) spin 1/2, corresponding to the $m^z_{J1} = + 1$ and $m^z_{J2} = + 2$ states. We construct the following effective spin-1/2 model Hamiltonian based on the idea of second-order perturbation theory, assuming that the two single-ion eigenstates, $|\!\!\downarrow\rangle\equiv|m^z_{J1} = +1\rangle$ and $|\!\!\uparrow\rangle\equiv|m^z_{J2} = +2\rangle$, are degenerate at a certain field between $H_{\rm m1}$ and $H_{\rm m2}$ (see Appendix~\ref{Appendix_Model}):
\begin{eqnarray}
\mathcal{H}_{\rm eff}^{\rm spin\text{-}1/2} &= &\mathcal{J}\sum_{\langle i,j\rangle}^{\rm plane}p_i^zp_j^z +\mathcal{J}^\prime\sum_{\langle\langle i,k\rangle\rangle}^{\rm plane}p_i^zp_k^z+\mathcal{W}\sum_{i,j,k\in \triangle,\triangledown}^{\rm plane}p_i^zp_j^zp_k^z\nonumber \\
&& - \tilde{g}\mu_{\rm B}\tilde{H}\sum_i p_i^z,
\label{etoh_yamamoto_model}
\end{eqnarray}
where $p^z|\!\!\uparrow\rangle =1/2|\!\!\uparrow\rangle$ and $p^z|\!\!\downarrow\rangle =-1/2|\!\!\downarrow\rangle$. 
Here, $\mathcal{J}$ and $\mathcal{J}^\prime$ denote the nearest-neighbor and next-nearest-neighbor interactions and $\mathcal{W}$ represents the three-body interactions among the three spins forming a regular triangle, as shown in Fig.~\ref{fig_interactions}(a). We assume only Ising-type couplings due to the observed step-like magnetization curves and omit the interlayer ferromagnetic exchanges since they contribute only a constant shift to the ground-state energy for Ising spins. 

\begin{figure}[t]
\begin{center}
\includegraphics[keepaspectratio, scale=0.7]{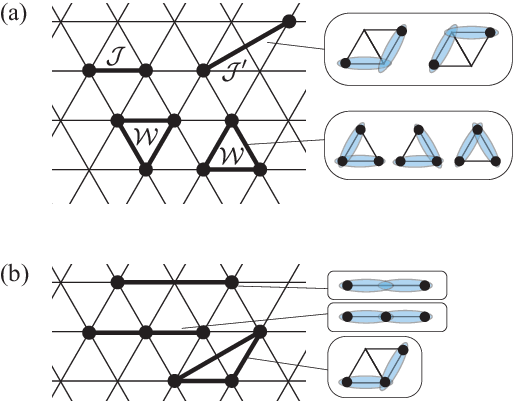}
\caption{(a) Schematic illustrations of the nearest-neighbor, next-nearest-neighbor, and three-body interactions considered in the present effective model. (b) Third-nearest-neighbor interaction and other types of three-body interactions, omitted for simplicity. Second-order processes contributing to each interaction are schematically shown as well (ovals represent the perturbative nearest-neighbor couplings).}\label{fig_interactions}
\end{center}
\end{figure}

\begin{figure*}[t]
\begin{center}
\includegraphics[keepaspectratio, scale=0.41]{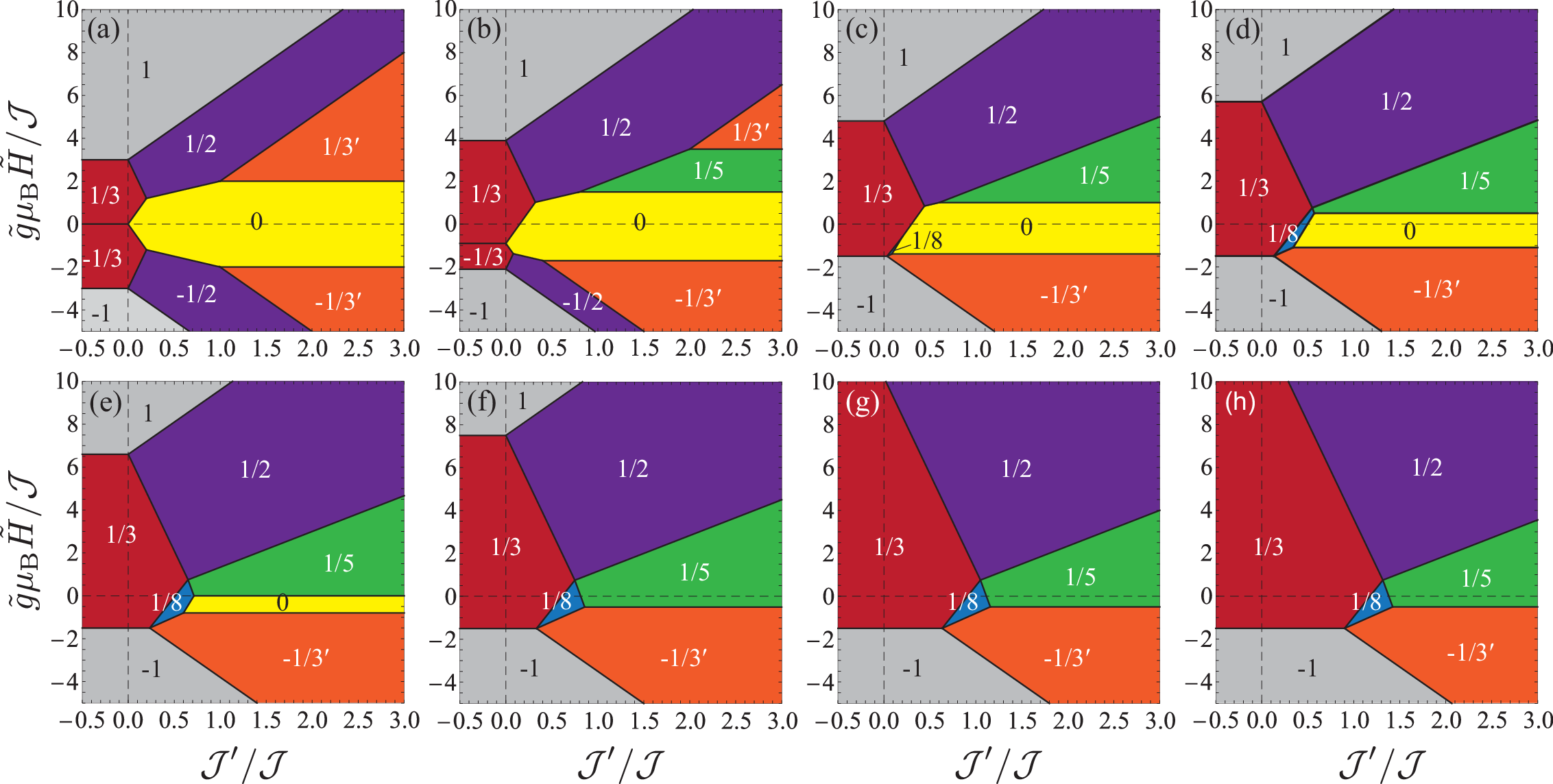}
\caption{(a-h) Ground-state phase diagrams of the effective $S=1/2$ model for the high-field metamagnetic transitions as functions of the shifted magnetic field $\tilde{g}\mu_B\tilde{H}$ ($\tilde{H}=H-H_0$) and the next-nearest-neighbor interaction $\mathcal{J}^\prime$, expressed as ratios to the nearest-neighbor interaction $\mathcal{J}$. We assume the presence of a three-body interaction acting on unit triangles, with several values of $\mathcal{W}/\mathcal{J}$: (a) 0.0, (b) 0.6, (c) 1.2, (d) 1.8, (e) 2.4, (f) 3.0, (g) 4.8, and (h) 6.4.}\label{fig_PDs}
\end{center}
\end{figure*}

{The magnetic field in Eq.~\eqref{etoh_yamamoto_model}, $\tilde{H}\equiv H-H_0$,} is defined relative to $H_0$ ($H_{\rm m1}<H_0<H_{\rm m2}$), which corresponds to the zero-field point of the effective model. The value of $H_0$ is later determined by fitting to the experimental data, together with the parameters $\mathcal{J}$, $\mathcal{J}^\prime$ , and $\mathcal{W}$. The $g$-factor $\tilde{g}$ is not necessarily identical to that in the effective spin-1 model at low fields. {A magnetic field $H > H_0$ corresponds to a positive field $\tilde{H}>0$, while $H < H_0$ represents a negative field $\tilde{H}<0$ in the language of the model~\eqref{etoh_yamamoto_model}. When a strong field is applied, the system enters a fully polarized state within the manifold of the effective model: for $\tilde{H}\gg 0$, all spins align in the $|\!\!\uparrow\rangle \equiv |m^z_{J2} = +2\rangle$ state, whereas for $\tilde{H} \ll 0$, all spins align in the $|\!\!\downarrow\rangle \equiv |m^z_{J1} = +1\rangle$ state. We aim to describe the observed metamagnetic transitions at high fields as magnetization processes between these two opposite saturation states, in the language of the effective spin-1/2 model. }

{In contrast to the conventional approach starting from degenerate low-energy states at zero field~\cite{Shiba2003perturbation}, the effective model derived from high-field states is no longer constrained by time-reversal symmetry. This crucial difference allows a three-body interaction term $\mathcal{W}$ to emerge, which would otherwise be forbidden in time-reversal symmetric settings.} Typically, the single-ion eigenstates selected as unperturbed states consist of time-reversal pairs (Kramers doublet), e.g., $|m^z_{J1}=\pm 1\rangle$, which leads to the cancellation of terms with an odd number of spin operators in the final form of the effective Hamiltonian~\cite{Shiba2003perturbation}. However, in this case, the two unperturbed states not only have different $|m^{z}|$ values but even belong to different $J$ manifolds, preventing such a cancellation. {See Appendix~\ref{Appendix_Model} for more details.}

\begin{figure}[t]
\begin{center}
\includegraphics[keepaspectratio, scale=0.28]{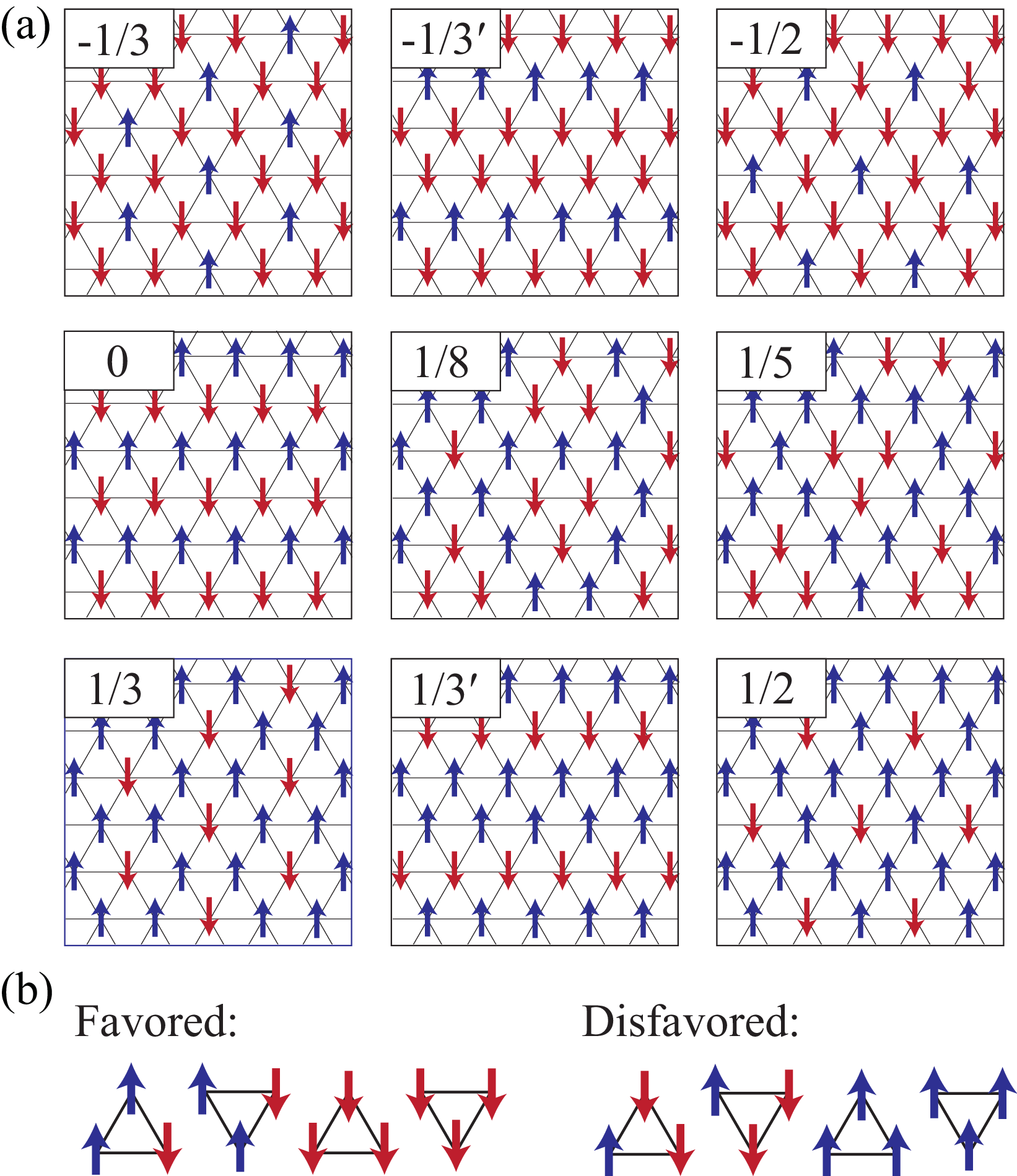}
\caption{(a) Schematic illustrations of the spin configurations for each state appearing in the phase diagrams shown in Figs.~\ref{fig_PDs}(a)-\ref{fig_PDs}(h). (b) Favored and disfavored spin configurations in a unit triangle when $\mathcal{W}>0$. }\label{fig_states}
\end{center}
\end{figure}

Note that second-order processes of the superexchange interactions via the original $S=2$ spins on neighboring Fe$^{2+}$ ions give rise to third-nearest-neighbor two-body couplings and other types of three-body interaction (``isosceles triangle'' and ``line''), as shown in Fig.~\ref{fig_interactions}(b). However, based on the counting of second-order processes that generate each term, the coefficient of the third-nearest-neighbor coupling is half that of the next-nearest-neighbor one, while the coefficient of the isosceles-triangle or line-shaped three-body coupling is one-third that of the regular triangle coupling [see Figs.~\ref{fig_interactions}(a) and~\ref{fig_interactions}(b)]. Thus, these terms are omitted here to construct a minimal model that captures the essential features of the observed metamagnetic transitions.

To begin, we first investigate the ground-state phase diagram of the effective spin-1/2 model given by Eq.~(\ref{etoh_yamamoto_model}) before fitting it to the experimental data. Figures~\ref{fig_PDs}(a)-\ref{fig_PDs}(h) show the ground-state phase diagrams in the 
$\tilde{g}\mu_B\tilde{H}/\mathcal{J}$ vs. $\mathcal{J}^\prime/\mathcal{J}$ plane with $\mathcal{J}>0$ for several values of $\mathcal{W}/\mathcal{J}\geq 0$. To obtain these phase diagrams, we generated all distinct  spin configurations on $L_x\times L_y$ parallelogram clusters of several different sizes, specifically $(L_x,L_y)=(3,4), (3,5), (3,6), (4,4), (4,5), (4,6), (5,5)$, and determined the one with the lowest energy per site under periodic boundary conditions for each given set of Hamiltonian parameters. We identified nine distinct spin configurations appearing in the ground-state phase diagram, depicted in Fig.~\ref{fig_states}(a), in addition to the fully polarized ``down'' and ``up'' states. Those states are characterized by their magnetization per saturation value, $ \tilde{M}\equiv \sum_i 2p_i^z / L_xL_y~(-1\leq  \tilde{M}\leq 1)$, with the caveat that two distinct states exist for $\tilde{M}=\pm 1/3$, respectively, which we differentiate using a prime. Note that the results for $\mathcal{W}<0$ can be trivially obtained by simultaneously applying the transformations $\tilde{H}\rightarrow -\tilde{H}$, $\mathcal{W}\rightarrow -\mathcal{W}$, and $
\tilde{M} \rightarrow -\tilde{M}$, due to the symmetry of the model. 

Let us now discuss the physical origin of the relationship between the model parameters and the ground-state spin configurations. The energy per site $E_{\tilde{M}}$ for each state with the scaled magnetization $\tilde{M}$ is given by 
\begin{equation}
\begin{aligned}
    E_{\pm 1} &= \mp \frac{1}{2}\tilde{g}\mu_B{\tilde{H}} + \frac{3}{4} \mathcal{J} + \frac{3}{4} \mathcal{J}^\prime \pm \frac{1}{4}\mathcal{W} \\
    E_{\pm 1/2} &= \mp \frac{1}{4}\tilde{g}\mu_B{\tilde{H}} \mp \frac{1}{8}\mathcal{W}  \\
    E_{\pm 1/3} &= \mp \frac{1}{6}\tilde{g}\mu_B{\tilde{H}} - \frac{1}{4}\mathcal{J} + \frac{3}{4} \mathcal{J}^\prime  \mp \frac{1}{4}\mathcal{W}  \\
    E_{\pm 1/3^\prime} &= \mp\frac{1}{6} \tilde{g}\mu_B{\tilde{H}} + \frac{1}{12}\mathcal{J} - \frac{1}{4}\mathcal{J}^\prime \pm \frac{1}{12}\mathcal{W} \\
    E_{\pm 1/5} &= \mp\frac{1}{10}\tilde{g}\mu_B{\tilde{H}} - \frac{1}{20}\mathcal{J} - \frac{1}{4}\mathcal{J}^\prime \mp \frac{1}{12}\mathcal{W} \\
    E_{\pm 1/8} &= \mp\frac{1}{16}\tilde{g}\mu_B{\tilde{H}} - \frac{3}{16} \mathcal{J} + \frac{3}{16} \mathcal{J}^\prime \mp \frac{5}{32} \mathcal{W} \\
    E_{0} &= -\frac{1}{4}\mathcal{J} - \frac{1}{4}\mathcal{J}^\prime 
\end{aligned}\label{ene_plateaus}
\end{equation}

\begin{figure}[t]
\begin{center}
\includegraphics[keepaspectratio, scale=0.5]{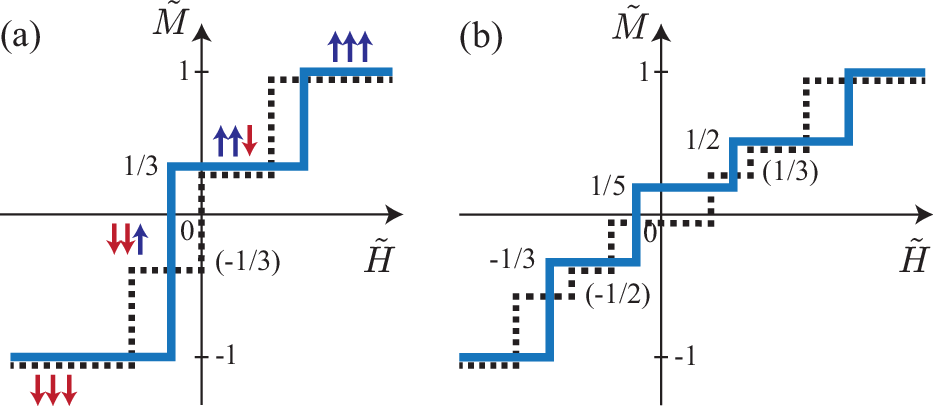}
\caption{(a,b) Schematic plots of the theoretically expected magnetization per saturation value, $\tilde{M}$, as a function of the shifted magnetic field $\tilde{H}=H-H_0$ (blue solid lines), for (a) $\mathcal{W}/\mathcal{J} \gtrsim 1$ and $\mathcal{J}^\prime/\mathcal{J} < 0$ (low-pressure regime) and (b) $\mathcal{W}/\mathcal{J} \gg 1$ and $\mathcal{J}^\prime/\mathcal{J} \gtrsim 1$ (high-pressure regime). For comparison, the corresponding curves in the absence of the three-body interaction ($\mathcal{W} = 0$) are also shown as dashed lines in each panel, slightly shifted downward for better visibility. The $\mathcal{W}=0$ curves display antisymmetry (odd-function symmetry) with respect to the shifted magnetic field, whereas finite $\mathcal{W}$ breaks this property and yields asymmetric profiles.}
\label{fig_magnetization_curve}
\end{center}
\end{figure}

{Note that the (-1/8) state and (-1/5) state do not appear as ground states in the current parameter range, but their energies are included for completeness.}
The $(\pm 1/3)$ states are the most typical spin structure observed as magnetization plateaus in triangular-lattice antiferromagnets~\cite{Nishimori1986-gr,Chubukov1991-gp,Ono2003-qf,Shirata2012-ej,Starykh2015-oj}. The $1/3$ state and the $(-1/3)$ state are time-reversal (spin-reversal) counterparts that are usually equivalent and appear for opposite directions of the magnetic field. From Eq.~\eqref{ene_plateaus}, it can be seen that a negative (ferromagnetic) next-nearest-neighbor interaction $\mathcal{J}^\prime<0$ further stabilizes the ($\pm 1/3)$ states, since all next-nearest-neighbor spin pairs align in the same direction, as illustrated in Fig.\ref{fig_states}(a). Of particular interest is that the three-body interaction $\mathcal{W}>0$ selectively favors the $1/3$ state while disfavoring the $(-1/3)$ state. This is because, in the $1/3$ state [$(-1/3)$ state], all the unit triangles contain two up and one down (two down and one up) spins. The three-body term in Eq.~\eqref{etoh_yamamoto_model} lowers the system's energy when the three spins $p^z_ip^z_jp^z_k$ on a unit triangle have an odd number of down spins for $W>0$ [see Fig.~\ref{fig_states}(b)]. Consequently,  for $\mathcal{J}^\prime <0$, the magnetization curve exhibits an imbalance in the plateau widths: the $1/3$-plateau becomes wider, while the $(-1/3)$-plateau becomes narrower; for sufficiently large $\mathcal{W}$, the $(-1/3)$-plateau may even disappear. Such an asymmetry between the $(\pm 1/3)$-plateaus under opposite field directions has not been observed in conventional triangular-lattice antiferromagnets~\cite{Nishimori1986-gr,Chubukov1991-gp,Ono2003-qf,Shirata2012-ej,Starykh2015-oj} without three-body interactions. 

For $\mathcal{J}^\prime >0$, the ground states become more diverse due to the competition between the three-body interaction and the antiferromagnetic next-nearest-neighbor interactions. Despite the simplicity of the model given in Eq.~\eqref{etoh_yamamoto_model}, states with complex sublattice structures, such as the $1/5$ and $1/8$ states, can emerge. These states exhibit characteristic spin structures, where clusters of down spins form unit-triangle shapes, favored by the three-body interaction $\mathcal{W}>0$. Additionally, there are other types of states with $\tilde{M}=\pm 1/3$, referred to as $(\pm 1/3^\prime)$ states, which differ from the previously mentioned $(\pm 1/3)$ states in that they have stripe-like patterns of up and down spins. These spin configurations can be stabilized even for large $\mathcal{J}^\prime >0$, and $\mathcal{W}>0$ favors the $(-1/3^\prime)$ state, in contrast to the case of the $(\pm 1/3)$ states. 

In the experimental observations shown in Fig.~\ref{Integral_f}, a single plateau appears between $H_{\rm m1}$ and $H_{\rm m2}$ at low pressures ($P\lesssim 1$ GPa). According to the theoretical phase diagrams shown in Fig.~\ref{fig_PDs}, this may correspond to a magnetization process from the $(-1)$ state to the $1/3$ state and then to the $1$ state, which is expected for $\mathcal{W}/\mathcal{J}\gtrsim 1$ and $\mathcal{J}^\prime<0$. Note that, in the experimental plots, the magnetization in Fig.~\ref{Integral_f}(a) and the integrated PDO signal in Figs.~\ref{Integral_f}(b)-\ref{Integral_f}(f)} are normalized such that 
the total jump associated with the metamagnetic transition is set to unity, 
and intermediate plateaus are labeled as fractions such as 1/3 or 2/3. 
In the theoretical description based on the effective spin-1/2 model, 
the magnetization per saturation value, $\tilde{M}$, is defined between $-1$ and $1$. 
Therefore, the two normalizations are related by ${\rm integrated}~|\Delta f_{\rm sub}| = (\tilde{M}+ 1)/2$ (or ${\rm normalized}~M = (\tilde{M}+ 1)/2$) so that the $2/3$ plateau observed experimentally corresponds to 
the $1/3$ plateau in the theoretical scale. At high pressures ($P\gtrsim 1$ GPa), three plateaus seem to emerge, which may correspond to a sequence of the $(-1/3^\prime)$ state, the $1/5$ state, and the $1/2$ state, expected for $\mathcal{W}/\mathcal{J}\gg 1$ and $\mathcal{J}^\prime/\mathcal{J}\gtrsim 1$. Therefore, within the framework of the effective spin-1/2 model in Eq.~\eqref{etoh_yamamoto_model}, the experimentally observed pressure dependence of the metamagnetic transition process can be interpreted as the result of an increase in $\mathcal{J}^\prime/\mathcal{J}$ from a negative to positive value, while maintaining $\mathcal{W}/\mathcal{J}$ as a large positive value. The theoretically expected change in the metamagnetic transition process from low to high pressures is sketched in Figs.~\ref{fig_magnetization_curve}(a) and~\ref{fig_magnetization_curve}(b).

To quantify the pressure dependence of the model parameters, we perform a least-squares fit to the experimental transition fields presented in Fig.~\ref{phase_diagram}(b). Specifically, we analyze the field separations measured from \( H_{\rm m1} \) at both low and high pressures, using distinct sets of theoretical expressions. 
At low pressures (\( P \leq 0.83\,\mathrm{GPa} \)), the transitions at \( H_{\rm m1} \) and \( H_{\rm m2} \) are interpreted as those between the (-1) state and the 1/3 state, and between the 1/3 state and the 1 state, respectively. Recalling that $\tilde{H} = H - H_0$, the corresponding theoretical expressions are given by
\begin{eqnarray}
  H_{\rm m1} = -\frac{3 \mathcal{J}}{2\tilde{g}\mu_B} + H_0,~H_{\rm m2} = \frac{6 \mathcal{J} + 3 \mathcal{W}}{2\tilde{g}\mu_B} + H_0,
\end{eqnarray}
which leads to a simple relation for the plateau width:
\begin{equation}
  H_{\rm m2} - H_{\rm m1} = \frac{9 \mathcal{J} + 3 \mathcal{W}}{2\tilde{g}\mu_B}.
  \label{lowHH}
\end{equation}

At higher pressures (\( P \geq 1.23\,\mathrm{GPa} \)), the observed transition fields \( H_{\rm m1} \), \( H_{\rm a} \), \( H_{\rm b} \), and \( H_{\rm m2} \) are associated with a sequence of transitions between the (-1), (-1/3), 1/5, 1/2, and 1 states. The corresponding theoretical expressions are
\begin{eqnarray}
  H_{\rm m1} &=& -\frac{4 \mathcal{J} + 6 \mathcal{J}^\prime - \mathcal{W}}{2\tilde{g}\mu_B} + H_0, \\  
  H_{\rm a}   &=& -\frac{\mathcal{J}}{2\tilde{g}\mu_B} + H_0, \\  
  H_{\rm b}   &=& \frac{6\mathcal{J} + 30\mathcal{J}^\prime - 5\mathcal{W}}{18\tilde{g}\mu_B} + H_0, \\  
  H_{\rm m2} &=& \frac{6 \mathcal{J} + 6 \mathcal{J}^\prime + 3 \mathcal{W}}{2\tilde{g}\mu_B} + H_0,
\end{eqnarray}
from which we obtain the field separations:
\begin{eqnarray}
  H_{\rm a} - H_{\rm m1} &=& \frac{3 \mathcal{J} + 6 \mathcal{J}^\prime - \mathcal{W}}{2\tilde{g}\mu_B}, \label{highHH1} \\
  H_{\rm b} - H_{\rm m1} &=& \frac{21 \mathcal{J} + 42 \mathcal{J}^\prime - 7 \mathcal{W}}{9\tilde{g}\mu_B}, \label{highHH2} \\
  H_{\rm m2} - H_{\rm m1} &=& \frac{5 \mathcal{J} + 6 \mathcal{J}^\prime + \mathcal{W}}{\tilde{g}\mu_B}. \label{highHH3}
\end{eqnarray}

By fitting these expressions to the experimental data, we extract the pressure dependence of the model parameters. Within the uncertainty of the fit, $\mathcal{J}$ and $\mathcal{W}$ are essentially pressure independent, whereas $\mathcal{J}^\prime$ exhibits a clear linear dependence on $P$. The best-fit forms are
\begin{equation}
\begin{aligned}
\frac{\mathcal{J}(P)}{\tilde{g}\mu_B} &= 0.0645 \quad \text{[T]}, \\
\frac{\mathcal{J}^\prime(P)}{\tilde{g}\mu_B} &= -0.0959+0.116\,P \quad \text{[T]}, \\
\frac{\mathcal{W}(P)}{\tilde{g}\mu_B} &= 0.196 \quad \text{[T]},
\end{aligned}
\label{parameters}
\end{equation}
where pressure \( P \) is measured in gigapascals (GPa).
In terms of relative scales, $\mathcal{W}/\mathcal{J}\!\approx\!3.04$, showing that $\mathcal{W}$ is several times larger than $\mathcal{J}$. By contrast, $\mathcal{J}^\prime(P)/\mathcal{J}$ varies strongly with pressure, changing sign at $P \approx 0.8$ $\mathrm{GPa}$ from negative to positive values.

\begin{figure}[t]
\begin{center}
\includegraphics[keepaspectratio, scale=0.64]{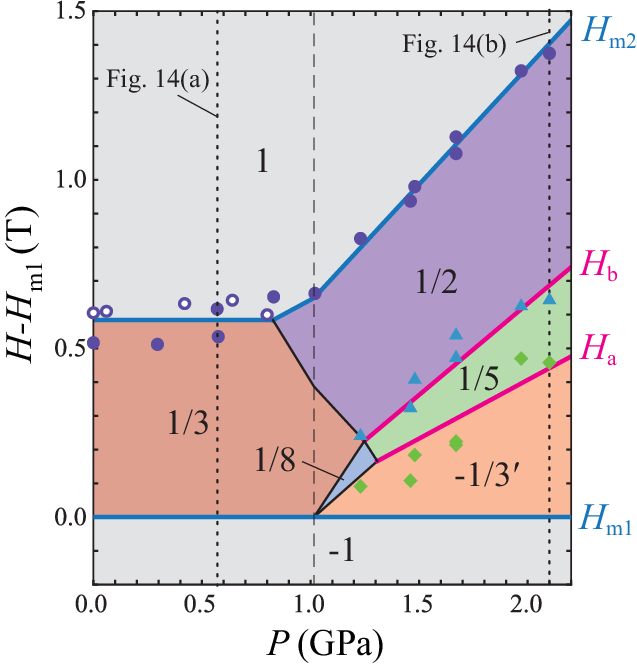}
\caption{Theoretical ground-state phase diagram as a function of magnetic field \( H \) (measured relative to \( H_{\rm m1} \)) and pressure \( P \), calculated using the fitted interaction parameters \( \mathcal{J}(P) \), \( \mathcal{J}^\prime(P) \), and \( \mathcal{W}(P) \) shown in Eq.~\eqref{parameters}. A discontinuity appears in the phase boundaries across \( P = P_{\rm c}^{\rm (th)} \equiv 1.02\,\mathrm{GPa} \) (indicated by the vertical dashed line), reflecting a change in the reference transition: For \( P < P_{\rm c}^{\rm (th)} \), \( H_{\rm m1} \) corresponds to the transition from the (-1) state to the 1/3 state, while for \( P > P_{\rm c}^{\rm (th)} \), it marks the transition from the (-1) state to the \( (-1/3^\prime) \) state. Experimental transition fields in the metamagnetic regime are overlaid for comparison, using the same symbols as in Fig.~\ref{phase_diagram}(b). The vertical dotted lines indicate the slices corresponding to Figs.~\ref{fig_comparison}(a) and~\ref{fig_comparison}(b).\label{theoretical_PD}}
\end{center}
\end{figure}

\begin{figure}[t]
\begin{center}
\includegraphics[keepaspectratio, scale=0.5]{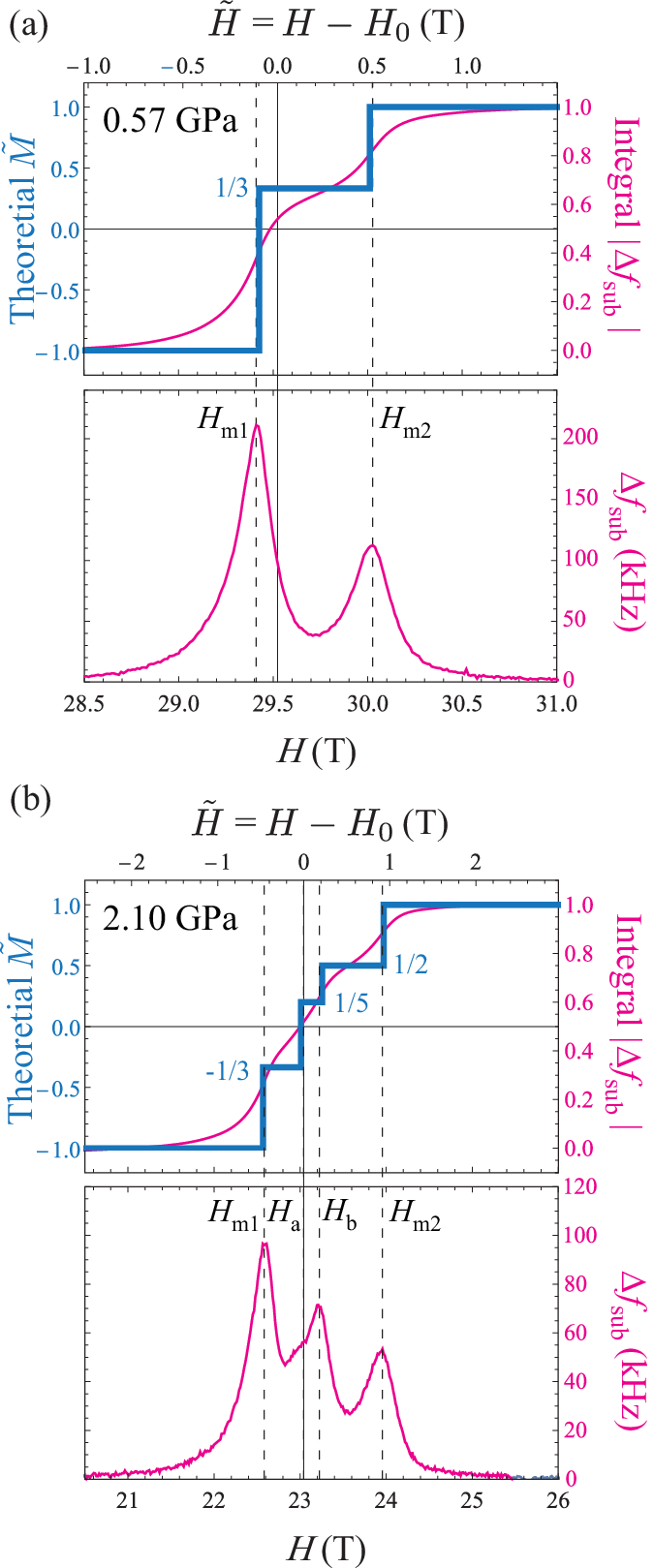}
\caption{Comparison between theory and experiment in the magnetization process for (a) $P = 0.57$ GPa and (b) $P = 2.10$ GPa. The center of the metamagnetic transition regime is set to (a) $H_0 = 29.52$ T and (b) $H_0=23.04$ T, respectively. }\label{fig_comparison}
\end{center}
\end{figure}

Based on the parameter functions in Eq.~\eqref{parameters}, we compare the theoretical ground-state phase diagram with the low-temperature experimental data in the plane of magnetic field and pressure, as shown in Fig.~\ref{theoretical_PD}. Here, we identify the onset of the metamagnetic transition, $H_{m1}$ as  the transition field from the (-1) state to the 1/3 state for $P < P_c^{\rm (th)} \equiv 1.02$ GPa, or to the $(-1/3^\prime)$ state for $P > P_c^{\rm (th)}$. Note that since the fitted value of $\mathcal{W}/\mathcal{J}$ is $\approx 3.04$, Fig.~\ref{theoretical_PD} can be regarded as approximately a replot of Fig.~\ref{fig_PDs}(f) (calculated with $\mathcal{W}/\mathcal{J}=3$), where the horizontal axis originally given by $\mathcal{J}'/\mathcal{J}$ is converted to pressure $P$ using Eq.~\eqref{parameters}, and the vertical axis is shifted by $H_{\mathrm{m}1}$.

Overall, the experimental data are very well explained by the theory. The theoretical prediction, however, suggests that the magnetization process could involve additional features not clearly resolved in the experiment, such as a possible transition from the 1/3 to the 1/2 state in an intermediate pressure regime around $P \sim 1$ GPa. One possible reason why this feature is not evident experimentally is that the change in the parameter ratio $\mathcal{J}^\prime(P)/\mathcal{J}(P)$ may occur more abruptly than the simple linear dependence assumed around $P \sim 1$ GPa. In addition, experimentally unavoidable finite-temperature effects and transverse spin-interaction components beyond the present Ising-type effective model may further affect the precise location of such a transition boundary. The theoretical phase diagram predicts the emergence of a 1/8-plateau. A signal that might correspond to this plateau is observed at $P \approx 1.23$ GPa, although more precise measurements are required to firmly establish its presence.

Additionally, in Fig.~\ref{fig_comparison}, we compare the theoretical predictions with experimental observations of the magnetization process at $P = 0.57$ and $P = 2.10$ GPa as representative examples. The origin of the magnetic field in the effective model is adjusted such that $H_0 = 29.52$ T and $23.04$ T, respectively, so that the center of the metamagnetic transition regime matches the experimental data. We find that the positions of the anomalies in the experimental $\Delta f_{\mathrm{sub}}$ are nicely explained by the theoretical calculation, by the predicted onset and termination fields of each plateau in the effective spin-1/2 model. The integrated $|\Delta f_{\mathrm{sub}}|$, although not strictly proportional to the magnetization, is expected to scale nearly with it, and its overall behavior is also captured well by the theory. The main difference is that the theoretical plateaus appear sharp, whereas the experimental curves are smoothened and more gradual. The observed broadening of the integrated signal may stem in part from the fact that it is not a direct probe of the magnetization, and in part from experimental conditions such as finite temperature or non-Ising spin interactions, both of which are beyond the scope of the present effective model. Finally, a more precise experimental determination of the fractional values of the magnetization plateaus is needed to establish their correspondence with the theoretical predictions.

For completeness, we also evaluated $H_0$ at other pressures in the same manner. Fitting these values as a function of $P$ yields an almost linear dependence, $H_0 \approx 32.65 - 4.93 P$ [T], indicating that the shift of the field origin in the effective spin-1/2 model, or the center of the metamagnetic transition, can be consistently described by a simple pressure dependence.

\color{black}
\section{Summary}
We have investigated pressure effects on the magnetism of CsFeCl$_3$ in magnetic fields.
The pressure dependence of the transition fields $H_{\rm c1}$ and $H_{\rm c2}$ in the low- and intermediate-field region can be interpreted by the dispersion relation of the $J$ = 1 state and pressure-dependent exchange interactions, exchange anisotropy, and single-ion anisotropy. Based on the energy levels of the Fe$^{2+}$ ion, the pressure dependence of the metamagnetic transition fields ($H_{\rm m1}$ and $H_{\rm m2}$) can be explained by the change of the single-ion anisotropy in the lowest $J$ = 1 and the excited $J$ = 2 state due to the larger trigonal distortion with increasing pressure. The metamagnetic transition was experimentally confirmed to be attributed to the level crossing between the $m^{z}_{J1}=+1$ and $m^{z}_{J2} = +2$ states at $\mbox{\boldmath $k$}$ = $\mbox{\boldmath $Q$}$ = (1/3, 1/3, 0). With increasing pressure above 1.23 GPa, we observed additional magnetic transitions between $H_{\rm m1}$ and $H_{\rm m2}$, which show some stepwise magnetization curve with the 1/3, 1/2, and 3/4 plateaus.  

A key to understanding the field--pressure evolution of CsFeCl$_3$ is to treat the low- and high-field regimes as distinct descriptions by different effective spin models. Low and intermediate fields are adequately captured by the established spin-1 description, which accounts for the singlet ground state and its response to moderate fields. In contrast, the high-field regime requires a qualitatively different construction. By projecting onto Zeeman-selected states from the $J=1$ and $J=2$ crystal-field multiplets that are not connected by time-reversal symmetry, we obtain an effective spin-1/2 framework that departs fundamentally from conventional approaches. Within this basis, symmetry permits three-body interactions on triangular plaquettes, which are strictly forbidden in the standard spin Hamiltonian. Importantly, such a three-body model does not obey the antisymmetry relation of the magnetization curve, $\tilde{M}(\tilde{H})=-\tilde{M}(-\tilde{H})$, and its phase diagram as a function of the next-nearest-neighbor ($\mathcal{J}^\prime$) and three-body ($\mathcal{W}$) couplings is remarkably rich even in the Ising-type limit. In particular, additional fractional magnetization plateaus such as $1/5$ and $1/8$, absent in conventional triangular-lattice spin models, emerge. These states exhibit characteristic spin structures, where clusters of down spins form unit-triangle motifs stabilized by the three-body interaction $\mathcal{W}>0$. This framework naturally explains the mechanism of the experimentally observed metamagnetic transition and its pressure evolution: In the regime, where $\mathcal{W}$ dominates over the nearest-neighbor interaction $\mathcal{J}$, a pressure-driven sign change of $\mathcal{J}^\prime$ transforms the stepwise high-field metamagnetism into an increasingly rich sequence of fractional plateaus.

The metamagnetic transition of CsFeCl$_3$ near 32~T had remained a long-standing puzzle since its discovery in the 1980s~\cite{CHIBA1987427,tsuboi1988magnetization}, defying consistent microscopic interpretation for nearly four decades. Our present framework provides a coherent microscopic understanding of this transition, offering a plausible resolution to its anomalous high-field behavior. Beyond resolving this anomaly, we establish a broader principle for quantum magnets in high magnetic fields: when crystal-field multiplets that are distinct at zero field are brought into level crossing by Zeeman splitting, the resulting low-energy spin models are no longer constrained to even-body couplings but may naturally include odd-body interactions. This mechanism may be relevant not only to structurally similar compounds such as RbFeCl$_3$~\cite{Amaya1988-pz,Stoppel2021-vw}, which also exhibits high-field metamagnetism, but more broadly to a wide class of systems where multiplet physics and high magnetic fields coexist. These findings open a new avenue for exploring exotic phases and unconventional spin interactions beyond the conventional paradigm of two-body or four-body physics, offering opportunities for both theoretical discovery and experimental realization in quantum magnetism.

\color{black}

\begin{acknowledgments}
This study was supported by a Sasakawa Scientific Research Grant from the Japan Science Society and JST, the establishment of university fellowships towards the creation of science technology innovation, Grant Number JPMJFS2125. We acknowledge support from the Deutsche Forschungsgemeinschaft (DFG) through SFB 1143 (Project No.\ 247310070) and the Deutsche Forschungsgemeinschaft through the W\"{u}rzburg-Dresden Cluster of Excellence on Complexity and Topology in Quantum Matter -- $ct.qmat$ (EXC 2147, project No. 390858490) and SFB 1143 as well as by HLD at HZDR, member of the European Magnetic Field Laboratory (EMFL). The work of D.Y. was supported by JSPS KAKENHI Grant Nos.~21H05185, 23K25830, 24K06890, and JST PRESTO Grant No.~JPMJPR245D,
\end{acknowledgments}

\appendix

\section{Derivation of the Effective Spin-1/2 Model from Second-Order Processes}
\label{Appendix_Model}

In this appendix, we outline the derivation of the effective spin-1/2 model introduced in Eq.~(11), which describes the metamagnetic transition of CsFeCl$_3$ between the states \( |\!\!\downarrow\rangle \equiv |m^z_{J1} = 1\rangle \) and \( |\!\!\uparrow\rangle \equiv |m^z_{J2} = 2\rangle \) for high fields. These two states belong to different total angular momentum manifolds, \( J=1 \) and \( J=2 \), and are, therefore, not connected by time-reversal symmetry. This crucial distinction allows for the emergence of odd-order terms in the effective Hamiltonian, including the three-body interaction discussed in the main text.

To derive the effective model, we consider virtual processes, where the superexchange interaction between neighboring Fe\(^{2+} \) ions couples the low-energy states \( |\!\!\downarrow\rangle \) and \( |\!\!\uparrow\rangle \) via intermediate excited crystal-field states. This mechanism is analogous to the approach used by Shiba et al.~\cite{Shiba2003perturbation} in their study of CsCoCl\(_3\), where second-order perturbation theory accounts for contributions from crystal-field-excited Kramers doublets. In their case, time-reversal symmetry constrains the effective low-energy Hamiltonian within a Kramers doublet, forbidding three-body interactions. However, in our high-field regime, such constraints are absent.

Following standard degenerate perturbation theory, we construct the effective Hamiltonian by projecting the full Hilbert space onto the two-level subspace spanned by \( |\!\!\downarrow\rangle \equiv |m^z_{J1} = 1\rangle \) and \( |\!\!\uparrow\rangle \equiv |m^z_{J2} = 2\rangle \). The unperturbed Hamiltonian is a sum of local single-ion terms for each Fe\(^{2+}\) ion, given by
\begin{equation}
\mathcal{H}_0 = -k\lambda\, \bm{l} \cdot \bm{S} - \delta\left((l^z)^2 - \frac{2}{3}\right) - \mu_{\rm B} \bar{H}(-k l^z + 2 S^z),
\label{unperturbed}
\end{equation}
where \( \bm{l} \) is the effective orbital angular momentum (\( l=1 \)), \( \bm{S} \) is the original spin operator (\( S=2 \)), \( \lambda \) is the spin-orbit coupling constant, \( \delta \) denotes the trigonal crystal-field distortion, and \( k \) is the orbital reduction factor. The field \( \bar{H} \) is defined such that the two states \( |\!\!\downarrow\rangle \) and \( |\!\!\uparrow\rangle \) are degenerate in energy.

The perturbation consists of the residual magnetic field \( H - \bar{H} \) and the isotropic superexchange interaction between neighboring Fe\(^{2+}\) ions. These perturbations allow for virtual transitions to excited crystal-field states, which in turn mediate effective interactions between the two low-energy states. The eigenstates \( |\!\!\downarrow\rangle \) and \( |\!\!\uparrow\rangle \) are obtained by diagonalizing \( \mathcal{H}_0 \) and are expressed as linear combinations of the basis states \( |l_z, S_z\rangle \) satisfying $m_l+ m_S=1$ and 2, respectively:
\begin{eqnarray}
|\!\!\downarrow\rangle &=& \sum_{m_l+ m_S=1} c^{(\downarrow)}_{m_l, m_S} |l^z = m_l, S^z = m_S\rangle, \\
|\!\!\uparrow\rangle &=& \sum_{m_l+ m_S=2} c^{(\uparrow)}_{m_l, m_S} |l^z = m_l, S^z = m_S\rangle. 
\end{eqnarray}
The matrix elements of the spin-2 operators $\bm{S}$ within the two-level subspace are given by 
\begin{eqnarray*}
\langle \uparrow|S^z|\uparrow\rangle &=& |c^{(\uparrow)}_{1, 1}|^2+2|c^{(\uparrow)}_{0, 2}|^2,\\
\langle \downarrow|S^z|\downarrow\rangle &=& |c^{(\downarrow)}_{0, 1}|^2+2|c^{(\downarrow)}_{-1, 2}|^2,\\
\langle \uparrow|S^+|\downarrow\rangle &=& \langle \downarrow|S^-|\uparrow\rangle^\ast=\sqrt{6}c^{(\uparrow)\ast}_{1, 1}c^{(\downarrow)}_{1, 0}+2c^{(\uparrow)\ast}_{0, 2}c^{(\downarrow)}_{0, 1}.
\end{eqnarray*}
All other combinations vanish due to angular-momentum selection rules. 

We now focus on the perturbation arising from the superexchange interactions between Fe$^{2+}$ ions located in the same triangular-lattice planes. These interactions couple nearest-neighbor spins via isotropic exchange of the form
\begin{equation}
V = I \sum_{\langle i,j\rangle}^{\rm plane} \bm{S}_i\cdot\bm{S}_j,
\end{equation}
where \( I > 0 \) denotes the antiferromagnetic exchange coupling between in-plane Fe$^{2+}$ ions. The effective first-order Hamiltonian is given by \( H^{(1)}_{\mathrm{eff}} = \mathcal{P} V \mathcal{P} \), where \( \mathcal{P} \) is the projection onto the subspace spanned by \( |\!\!\downarrow\rangle \) and \( |\!\!\uparrow\rangle \).

In this subspace, the effective Hamiltonian at first order generally contains both diagonal and off-diagonal components, leading to
\begin{equation}
\mathcal{H}^{(1)}_{\mathrm{eff}} = \sum_{\langle i,j \rangle}^{\text{plane}} \left(  \mathcal{J}_{\perp}^{(1)} (p^x_i p^x_j + p^y_i p^y_j)+\mathcal{J}_{\parallel}^{(1)} p^z_i p^z_j  \right) + \Delta^{(1)} \sum_i p^z_i,
\end{equation}
in terms of the effective pseudospin-1/2 operators \( \bm{p}_i \) acting on the two pseudospin states \( |\!\!\downarrow\rangle \equiv |m^z_{J1} = 1\rangle \) and \( |\!\!\uparrow\rangle \equiv |m^z_{J2} = 2\rangle \). The coupling constants \( \mathcal{J}_{\parallel}^{(1)} \) and \( \mathcal{J}_{\perp}^{(1)} \) are determined by the projected matrix elements of the spin-2 operators within the low-energy subspace. The transverse coupling is given by
\begin{equation}
\mathcal{J}_{\perp}^{(1)} = I\, |\langle \uparrow | S^+ | \downarrow \rangle|^2 = I \left| \sqrt{6}\, c^{(\uparrow)*}_{1,1} c^{(\downarrow)}_{1,0} + 2\, c^{(\uparrow)*}_{0,2} c^{(\downarrow)}_{0,1} \right|^2,
\end{equation}
while the longitudinal coupling is
\begin{eqnarray}
\mathcal{J}_{\parallel}^{(1)} &=& I \left( \langle \uparrow | S^z | \uparrow \rangle - \langle \downarrow | S^z | \downarrow \rangle \right)^2 \nonumber\\
&=& I \left( |c^{(\uparrow)}_{1,1}|^2 + 2|c^{(\uparrow)}_{0,2}|^2 - |c^{(\downarrow)}_{0,1}|^2 - 2|c^{(\downarrow)}_{-1,2}|^2 \right)^2.
\end{eqnarray}
The effective Zeeman term \( \Delta^{(1)} \) also appears at this order and is given by
\begin{eqnarray}
\Delta^{(1)} &=& \frac{z}{2}I \left( \langle \uparrow | S^z | \uparrow \rangle^2 - \langle \downarrow | S^z | \downarrow \rangle^2 \right) \nonumber\\
&=& \frac{z}{2}I \left[ \left( |c^{(\uparrow)}_{1,1}|^2 + 2|c^{(\uparrow)}_{0,2}|^2 \right)^2 - \left( |c^{(\downarrow)}_{0,1}|^2 + 2|c^{(\downarrow)}_{-1,2}|^2 \right)^2 \right],
\end{eqnarray}
where \( z = 6 \) is the coordination number of the triangular lattice. 

Unlike the case, where the low-energy doublet forms a symmetric Kramers pair protected by time-reversal symmetry~\cite{Shiba2003perturbation}, the present states are not related by time reversal and arise from different total angular momentum multiplets. Consequently, their matrix elements are not constrained to have equal magnitude:
\begin{equation}
|\langle \uparrow | S^z | \uparrow \rangle| \ne |\langle \downarrow | S^z | \downarrow \rangle|,
\end{equation}
which gives rise to a finite contribution to the residual field term \( H - \bar{H} \) in the effective spin-1/2 description.

We now move on to the second-order contributions arising from degenerate perturbation theory. The second-order effective Hamiltonian is given by~\cite{Mila2011}
\begin{equation}
\mathcal{H}^{(2)}_{\mathrm{eff}} = - \mathcal{P} V\mathcal{Q}  \frac{1}{\mathcal{Q}\mathcal{H}_0\mathcal{Q} - E_0}\mathcal{Q} V \mathcal{P},
\end{equation}
where $\mathcal{Q}\equiv 1-\mathcal{P}$ and \( E_0 \) refers to the common eigenvalue of the degenerate ground-state doublet under \( \mathcal{H}_0 \). Since the perturbation consists of nearest-neighbor (NN) superexchange interactions of the form \( \bm{S}_i \cdot \bm{S}_j \), each term in the second-order expression involves two such interactions, say \( \bm{S}_i \cdot \bm{S}_j \) and \( \bm{S}_k \cdot \bm{S}_l \), acting sequentially. Non-vanishing contributions to the effective Hamiltonian arise only when the two interaction terms share at least one site in common---that is, when the two bonds \( \langle i,j \rangle \) and \( \langle k,l \rangle \) are either identical or connected through a shared site. This constraint follows from the fact that in order to return to the low-energy subspace after a virtual excitation, the second interaction must act on at least one of the sites involved in the initial excitation.

Consequently, the second-order processes can be categorized as follows:
\begin{enumerate}
    \item {Same-bond processes}: When both interactions act on the same NN bond \( \langle i,j \rangle \), the virtual excitation and de-excitation occur on the same pair of sites. These processes yield further corrections to the two-body interactions between spins \( i \) and \( j \), and can also generate one-body shift of the Zeeman term.
    
    \item {Neighboring-bond processes}: When two NN bonds share a common site---i.e., \( \langle i,j \rangle \) and \( \langle j,k \rangle \)---the excitation created on one bond can be annihilated on the adjacent one. These processes give rise to effective three-body interactions involving three sites connected by two adjacent bonds, typically forming triangles or linear chains, such as \( p^z_i p^z_j p^z_k \), \( p^x_i p^z_j p^x_k + p^y_i p^z_j p^y_k \), or other scalar combinations consistent with lattice and spin symmetries. They can also produce next-nearest-neighbor two-body terms between sites \( i \) and \( k \), as well as an additional contribution to the Zeeman term.

    \item {Disconnected-bond processes}: When the two bonds do not overlap (i.e., \( \langle i,j \rangle \cap \langle k,l \rangle = \emptyset \)), the two interactions act on independent sites, and the resulting contribution cannot return to the low-energy subspace. Therefore, such processes do not contribute to \( H^{(2)}_{\mathrm{eff}} \).
\end{enumerate}
This classification provides a guiding framework for computing second-order contributions to the effective Hamiltonian and for interpreting their physical origin. 

Based on the classification of second-order virtual processes and the symmetries of the system, the resulting effective Hamiltonian can, in principle, include the following types of interactions among the projected spin-1/2 degrees of freedom. First, renormalizations of the NN two-body interactions appear as corrections to the longitudinal and transverse couplings: \( \mathcal{J}_\parallel^{(2)} p^z_i p^z_j \) and \( \mathcal{J}_\perp^{(2)} (p^x_i p^x_j + p^y_i p^y_j) \). In addition, next-nearest-neighbor two-body terms such as \( \mathcal{J}_\parallel^{\prime(2)} p^z_i p^z_k \) and \( \mathcal{J}_\perp^{\prime(2)} (p^x_i p^x_k + p^y_i p^y_k) \) can also emerge through second-order processes~\cite{Shiba2003perturbation}. Second, genuine three-body interactions involving three connected sites---such as \( p^z_i p^z_j p^z_k \) or symmetry-allowed combinations including \( p^x_i p^z_j p^x_k + p^y_i p^z_j p^y_k \)---can arise from neighboring-bond processes. Finally, one-body terms such as on-site energy shifts \( \Delta^{(2)} p^z_i \) may also appear, especially in the absence of symmetry protection between the two low-energy states.

In this work, we construct an effective spin-1/2 model, given in Eq.~\eqref{etoh_yamamoto_model}, to describe the high-field magnetic behavior of CsFeCl$_3$. Motivated by experimental observations indicating that the dominant features can be captured using only the longitudinal components of the spin interactions, we restrict the effective Hamiltonian to Ising-type couplings. Among the second-order terms, we further include only three-body interactions acting on equilateral triangular plaquettes of the triangular lattice, which are naturally expected to yield dominant contributions and help reduce the number of fitting parameters.

We note that the residual-field term \( H - \bar{H} \) is generally shifted by \( \Delta^{(1)} \), \( \Delta^{(2)} \), and higher-order corrections when time-reversal symmetry is broken. In Eq.~\eqref{etoh_yamamoto_model}, we express the Zeeman term as \( H - H_0 \), with \( H_0 \) effectively absorbing these interaction-induced shifts from the microscopic degeneracy point \( \bar{H} \).

Although all effective coupling constants can, in principle, be derived via perturbation theory using microscopic single-ion parameters of the spin-2 model---for example, \( \mathcal{J} = \mathcal{J}_\parallel^{(1)} + \mathcal{J}_\parallel^{(2)} \)---we instead determine them by directly fitting to experimental data. This approach is justified by the fact that the precise parameter regime in which the two relevant spin states become degenerate under strong magnetic fields remains controversial~\cite{CHIBA1987427}. Moreover, the fitted parameters often effectively incorporate higher-order corrections beyond the scope of low-order perturbation theory, leading to improved agreement with experimental observations.

We briefly remark on the distinction between the present mechanism and pseudospin-1/2 mappings commonly employed in strong-rung spin ladders and coupled dimer systems (see, e.g., Refs.~\cite{Tandon1999-qe,Furusaki1999-ps,Jaime2004-te}). While odd-body interactions are also symmetry-allowed in those settings due to the singlet-triplet basis, they typically arise only as subordinate corrections to dominant two-body terms and remain quantitatively minor. In contrast, the crystal-field origin of the effective spin-1/2 states in CsFeCl$_3$ allows for significantly stronger three-body couplings, which lead to pronounced field asymmetry around the zero point of the effective spin-1/2 model. (Having said that, singlet-triplet systems exhibit their own intriguing physics: transverse three-body corrections, interpreted as correlated hopping in the hardcore boson language~\cite{Picon2008-cg}, can give rise to exotic phases such as supersolidity~\cite{Picon2008-cg,Ng2006-gz,Yamamoto2013-wh}.)

\color{black}
\bibliography{nihongi_rsi}

\end{document}